\documentclass[twocolumn,twocolappendix]{aastex7}
\usepackage{natbib}

\usepackage{comment}
\usepackage{amsmath}
\usepackage{algorithm}
\usepackage{amsfonts}
\usepackage{mathrsfs}
\usepackage{amssymb}
\usepackage{color}
\usepackage{graphicx} 
\let\tablenum\relax
\usepackage{siunitx}

\newcommand{\cha}{\textit{Chandra} }

\newcommand{\nus}{\textit{NuSTAR} }
\newcommand{\xmm}{XMM-\textit{Newton} }

\newcommand{\jwst}{\textit{JWST} }
\newcommand{\thei}{3$-$8\,keV }
\newcommand{\thtw}{3$-$24\,keV }
\newcommand{\eitw}{8$-$24\,keV }
\newcommand{\eisi}{8$-$16\,keV }
\newcommand{\sitw}{16$-$24\,keV }
\newcommand{\zetw}{0.5$-$2\,keV }
\newcommand{\twte}{2$-$10\,keV }
\newcommand{\detml}{DET\_ML }
\newcommand{\logn}{log\textit{N}$-$log\textit{S} }

\begin{document}

\defcitealias{Zhao2021}{Z21}
\defcitealias{Zhao2024}{Z24}

\title{PEARLS: NuSTAR and XMM-Newton Extragalactic Survey of the JWST North Ecliptic Pole Time-domain Field III}

\correspondingauthor{Ross Silver}
\author[0000-0001-6564-0517]{Ross Silver}
\email{ross.m.silver@nasa.gov}
\affiliation{NASA Goddard Space Flight Center, Greenbelt, MD 20771, USA}
\affiliation{Southeastern Universities Research Association, Washington, DC 20005, USA}
\affiliation{School of Physics and Astronomy, University of Minnesota, Minneapolis, MN 55455, USA}

\author[0000-0002-2115-1137]{Francesca Civano} \email{francesca.m.civano@nasa.gov}
\affiliation{NASA Goddard Space Flight Center, Greenbelt, MD 20771, USA}

\author[0000-0002-7791-3671]{Xiurui Zhao} 
\email{xiurui.zhao.work@gmail.com}
\affiliation{Department of Astronomy, University of Illinois at Urbana-Champaign, Urbana, IL 61801, USA}
\affiliation{Cahill Center for Astrophysics, California Institute of Technology, 1216 East California Boulevard, Pasadena, CA 91125, USA}

\author[0000-0002-9041-7437]{Samantha Creech}
\email{s.creech@utah.edu}
\affiliation{Department of Physics \& Astronomy, University of Utah, 115 South 1400 East, Salt Lake City, UT 84112, USA}

\author[0000-0001-9262-9997]{Christopher N.\ A.\ Willmer} \email{cnawillmer@gmail.com}
\affiliation{Steward Observatory, University of Arizona,
933 N Cherry Ave, Tucson, AZ, 85721-0009, USA}

\author[0000-0002-9895-5758]{S.\ P.\ Willner} \email{swillner@cfa.harvard.edu}
\affiliation{Center for Astrophysics \textbar\ Harvard \& Smithsonian, 
60 Garden Street, Cambridge, MA, 02138, USA}

\author[0000-0001-8156-6281]{Rogier A.\ Windhorst} \email{Rogier.Windhorst@gmail.com}
\affiliation{School of Earth and Space Exploration, Arizona State University,
Tempe, AZ 85287-1404, USA}

\author[0000-0001-7592-7714]{Haojing Yan} 
\email{yanhaojing@gmail.com}
\affiliation{Department of Physics and Astronomy, University of Missouri,
Columbia, MO 65211, USA}

\author[0000-0002-6610-2048]{Anton M.\ Koekemoer}  \email{koekemoer@stsci.edu}
\affiliation{Space Telescope Science Institute,
3700 San Martin Drive, Baltimore, MD 21218, USA}

\author[0000-0003-3351-0878]{Rosalia O'Brien}
\email{robrien5@asu.edu}
\affiliation{School of Earth and Space Exploration, Arizona State University, Tempe, AZ 85287-1404, USA}

\author[0000-0002-6150-833X]{Rafael {Ortiz~III}}  \email{rortizii@asu.edu}
\affiliation{School of Earth and Space Exploration, Arizona State University,
Tempe, AZ 85287-1404, USA}

\author[0000-0003-1268-5230]{Rolf A.\ Jansen} \email{rolfjansen.work@gmail.com}
\affiliation{School of Earth and Space Exploration, Arizona State University,
Tempe, AZ 85287-1404, USA}

\author[0000-0002-2203-7889]{W. Peter Maksym} \email{walter.p.maksym@nasa.gov}
\affiliation{Center for Astrophysics \textbar\ Harvard \& Smithsonian, 
60 Garden Street, Cambridge, MA, 02138, USA}

\author[0000-0002-1697-186X]{Nico Cappelluti} \email{ncappelluti@miami.edu}
\affiliation{Department of Physics, University of Miami, Coral Gables, FL 33124, USA}

\author[0000-0002-9286-9963]{Francesca Fornasini} \email{ffornasini@stonehill.edu}
\affiliation{Stonehill College, 320 Washington Street, Easton, MA 02357, USA}

\author[0000-0001-6650-2853]{Timothy Carleton}  \email{tmcarlet@asu.edu}
\affiliation{School of Earth and Space Exploration, Arizona State University,
Tempe, AZ 85287-1404, USA}

\author[0000-0003-3329-1337]{Seth H.\ Cohen}  \email{seth.cohen@asu.edu}
\affiliation{School of Earth and Space Exploration, Arizona State University,
Tempe, AZ 85287-1404, USA}

\author[0000-0002-9984-4937]{Rachel Honor} 
\email{rchonor@asu.edu}
\affiliation{School of Earth and Space Exploration, Arizona State University, Tempe, AZ 85287-1404, USA}

\author[0000-0002-7265-7920]{Jake Summers} 
\email{jssumme1@asu.edu}
\affiliation{School of Earth and Space Exploration, Arizona State University,
Tempe, AZ 85287-1404, USA}

\author[0000-0002-9816-1931]{Jordan C.\ J.\ D'Silva}  \email{jordan.dsilva@research.uwa.edu.au}
\affiliation{International Centre for Radio Astronomy Research (ICRAR) and the
International Space Centre (ISC), The University of Western Australia, M468,
35 Stirling Highway, Crawley, WA 6009, Australia}
\affiliation{ARC Centre of Excellence for All Sky Astrophysics in 3 Dimensions
(ASTRO 3D), Australia}

\author[0000-0003-2714-0487]{Sibasish Laha} 
\email{sibasish.laha@nasa.gov}
\affiliation{Astrophysics Science Division, NASA Goddard Space Flight Center, Greenbelt, MD 20771, USA.}
\affiliation{Center for Space Science and Technology, University of Maryland Baltimore County, 1000 Hilltop Circle, Baltimore, MD 21250, USA.}
\affiliation{Center for Research and Exploration in Space Science and Technology, NASA/GSFC, Greenbelt, Maryland 20771, USA}

\author[0000-0001-7410-7669]{Dan Coe} 
\email{dcoe@stsci.edu}
\affiliation{Space Telescope Science Institute, 3700 San Martin Drive, Baltimore, MD 21218, USA}
\affiliation{Association of Universities for Research in Astronomy (AURA) for the European Space Agency (ESA), STScI, Baltimore, MD 21218, USA}
\affiliation{Center for Astrophysical Sciences, Department of Physics and Astronomy, The Johns Hopkins University, 3400 N Charles St. Baltimore, MD 21218, USA}

\author[0000-0003-1949-7638]{Christopher J.\ Conselice} \email{conselice@gmail.com}
\affiliation{Jodrell Bank Centre for Astrophysics, Alan Turing Building,
University of Manchester, Oxford Road, Manchester M13 9PL, UK}

\author[0000-0001-9065-3926]{Jose M. Diego}  \email{chemadiegor@gmail.com}
\affiliation{Instituto de F\'isica de Cantabria (CSIC-UC). Avenida. Los Castros
s/n. 39005 Santander, Spain}

\author[0000-0001-9491-7327]{Simon P.\ Driver} \email{Simon.Driver@icrar.org}
\affiliation{International Centre for Radio Astronomy Research (ICRAR) and the
International Space Centre (ISC), The University of Western Australia, M468,
35 Stirling Highway, Crawley, WA 6009, Australia}

\author[0000-0003-1625-8009]{Brenda Frye} 
\email{brendafrye@gmail.com}
\affiliation{Department of Astronomy/Steward Observatory, University of Arizona, 933 N Cherry Ave,
Tucson, AZ, 85721-0009, USA}

\author[0000-0001-9440-8872]{Norman A.\ Grogin}
\email{nagrogin@stsci.edu}
\affiliation{Space Telescope Science Institute,
3700 San Martin Drive, Baltimore, MD 21218, USA}

\author[0000-0001-6434-7845]{Madeline A.\ Marshall}  \email{madeline_marshall@outlook.com}
\affiliation{Los Alamos National Laboratory, Los Alamos, NM 87545, USA}

\author[0000-0003-3382-5941]{Nor Pirzkal} 
\email{npirzkal@stsci.edu}
\affiliation{Space Telescope Science Institute,
3700 San Martin Drive, Baltimore, MD 21218, USA}

\author[0000-0003-0429-3579]{Aaron Robotham} \email{aaron.robotham@uwa.edu.au}
\affiliation{International Centre for Radio Astronomy Research (ICRAR) and the
International Space Centre (ISC), The University of Western Australia, M468,
35 Stirling Highway, Crawley, WA 6009, Australia}

\author[0000-0003-0894-1588]{Russell E.\ Ryan, Jr.}  \email{rryan@stsci.edu}
\affiliation{Space Telescope Science Institute,
3700 San Martin Drive, Baltimore, MD 21218, USA}

\begin{abstract}
The \textit{James Webb Space Telescope} (JWST) North Ecliptic Pole (NEP) Time-Domain Field (TDF) has been monitored by \nus and \xmm with a regular cadence for five years starting in 2019. The survey has accumulated 3.5\,Ms of \nus exposure and 228\,ks quasi-simultaneous \xmm observations covering 0.31 deg$^2$. This paper presents the results from the most recent two-years' 2\,Ms \nus and 166\,ks XMM observations in \nus cycles 8 and 9. These observations reached a 20\%-area flux of 2.20 $\times$ 10$^{-14}$ erg cm$^{-2}$ s$^{-1}$ in the \eitw band. 75 \textit{\nus} sources and 274 \xmm sources are detected at 99\% reliability level. The logN$-$logS measured in cycles 8+9 are consistent with those measured in the previous cycle 5+6 NuSTAR NEP survey, but in a larger area (0.3 deg$^2$ compared with 0.19 deg$^2$). The slope of the cycles 8+9 \eitw \logn curve is flatter than other works ($\alpha_{89} = 1.13 \pm 0.46$), but is consistent with the Euclidean value of $\alpha = 1.50$. In addition, we found $\sim$36\% of the \nus sources to be heavily obscured ($N_{\rm H} \geq$ 10$^{23}$ cm$^{-2}$). The Compton-thick ($N_{\rm H} \geq$ 10$^{24}$ cm$^{-2}$) (CT-) AGN fraction is 9$^{+18}_{-8}$\% in the NEP-TDF, which is consistent with the measurements in previous surveys. %As the primary goal of this survey was time-domain science, we found 10 sources to be variable at a $\geq$2$\sigma$ level in at least one of the followings bands: \thei, \thtw, and \eitw.
\end{abstract}


\section{Introduction}
Active Galactic Nuclei (AGN) are supermassive black holes (SMBH) in the center of galaxies that accrete surrounding material and emit light across the entire electromagnetic spectrum. Numerous studies have found a strong correlation between the mass of the SMBH and the host galaxy bulge, host galaxy luminosity, and velocity dispersion \citep{Magorrian1998, Richstone1998, Gebhardt2000, Merritt2001, Ferrarese2005, Kormendy2013}. These correlations indicate that SMBHs and their host galaxies co-evolve with time and influence each other \citep{Fiore2017, Martin-Navarro2018}. Therefore, in order to understand galaxy evolution, we need a complete sample of AGN across cosmic time. \\
\indent One way to study AGN through time is by analyzing the Cosmic X-ray Background (CXB), i.e., the diffuse X-ray emission from 1 to $\sim$200$-$300\,keV encompassing the entire sky \citep{Gilli2007, Brandt2021CXB}, which is dominated by AGN. In particular, it is best to study the CXB at its peak \citep[20$-$40\,keV;][]{Ajello_2008} which is believed to be generated in large part ($\sim$10-50\%) by a subclass of obscured AGN known as Compton-thick AGN (CT-AGN) with neutral hydrogen column densities $\geq$10$^{24}$\,cm$^{-2}$ \citep{Maccacaro1988, Boyle1993, Comastri1995, Jones1997, Page1997, Boyle1998, Miyaji2000, Gilli2001, Cowie2003, Ueda2003, Gilli2007, Draper2009, Treister2009, Ueda2014, Aird2015, Buchner2015, Ananna2019}. Due to the extreme obscuration in these sources, much of the soft X-ray emission is suppressed. Therefore, only an instrument sensitive above 10\,keV can fully analyze CT-AGN and resolve the CXB completely. \\
\indent Launched in June of 2012, the Nuclear Spectroscopic Telescope Array (\textit{NuSTAR}) became the first telescope capable of focusing hard X-rays with a bandpass covering the 3$-$79\,keV range \citep{Harrison2013}. Compared with other collimated or coded-mask X-ray instruments, \nus is more sensitive by a factor of 10--100. Previous \nus extragalactic surveys have resolved $\sim$35\% of the peak of the CXB \citep{Harrison2016}. \\
\indent While several surveys were conducted in the first few years after the \nus launch -- \textit{Chandra} Deep Field-North survey \citep[CDFN, 0.07\,deg$^2$;][]{Alexander2003}, the Extended Groth Strip survey (EGS, 0.25\,deg$^2$; Aird et al. in preparation), the Extended \cha Deep Field-South \citep[ECDFS, 0.33\,deg$^2$;][]{Mullaney2015}, the Ultra Deep Survey (UDS) of the UKIRT Infrared Deep Sky Survey \citep[UKIDSS, 0.6\,deg$^2$;][]{Masini2018a}, and the Cosmic Evolution Survey Field \citep[COSMOS, 1.7\,deg$^2$;][]{Civano2015} -- none of them were designed to analyze time domain and variability. \\
\indent Our team has been granted four different Large category proposals from \nus cycles 5, 6, 8, and 9 (totaling $\sim$3.5\,Ms) to study the North Ecliptic Pole (NEP) Time-Domain Field \citep[TDF;][]{Jansen2018}. This field was selected as a region of interest for the James Webb Space Telescope \citep[\textit{JWST};][]{Gardner2006} for multiple reasons: 1) it is located in \textit{JWST's} northern continuous viewing zone (CVZ), making it available for observations all year round; 2) low Galactic foreground extinction; and 3) an absence of bright foreground stars (AB $\leq$ 16 mag). These three factors give this field a high potential for impactful time domain and population studies. As such, the Prime Extragalactic Areas for Reionization and Lensing Science (PEARLS) team \citep{Windhorst2023} was allocated $\sim$47 hours of guaranteed time to observe this field during cycle 1 (PI: R. Windhorst; program \textit{JWST}-GTO-1176). In addition to the \jwst data, this field has been studied across the electromagnetic spectrum in previous years\footnote{The complete table can be found in \url{http://lambda.la.asu.edu/jwst/neptdf/}}. It has been observed by: \textit{Chandra} (PI: Maksym), \textit{GTC}/HiPERCAM (PI: Dhillon), \textit{HST}/WFC3 + ACS (PI: Jansen), \textit{IRAM}/NIKA 2 (PI: Cohen), \textit{JCMT} (PI: Smail \& Im), \textit{J-PAS} (PI: Bonoli \& Dupke), \textit{LBT}/LBC (PI: Jansen), \textit{LOFAR} (PI: Van Weeren), \textit{MMT}/MMIRS + Binospec (PI: Willmer), \textit{Subaru}/HSC (PI: Hasinger \& Hu), \textit{TESS} (PI: Berriman \& Holwerda), \textit{VLA} (PI: Windhorst \& Cotton), and \textit{VLBA} (PI: Brisken). \\
\indent This extensive multi-wavelength coverage has enabled the combination of the infrared and hard X-rays, which has proven to be extremely effective in studying AGN, and in particular, obscured AGN, as these two bands are the least susceptible to obscuration. As a result, these wavelengths have been used to identify CT-AGN \citep[e.g.,][]{Padovani2017} and to study the host-galaxy properties of obscured AGN \citep[e.g.,][]{Gandhi2009}. Moreover, the \jwst data, despite covering a very small area, have provided accurate redshifts for some of the hard X-ray detected sources (see Section \ref{sec:redshifts}). \\
\indent As mentioned previously, our team has been granted time in four different \nus cycles covering five contiguous years to study the NEP-TDF. The results from the cycle 5 observations have been published in \cite{Zhao2021}, while the combined cycles 5 and 6 results were published in \cite{Zhao2024}. Going forward, these works will be referred to as \citetalias{Zhao2021} and \citetalias{Zhao2024}, respectively. These initial observations focused on depth rather than area, covering 0.16\,deg$^2$, but reaching a flux level of 1.7$\times$10$^{-14}$ erg cm$^{-2}$ s$^{-1}$ in the \eitw band. In this work, we report the results of the combined cycles 8+9 \nus and \xmm data which were acquired recently focusing on wider coverage (0.31\,deg$^2$), thus doubling the area covered by the previous survey. \\
\indent As discussed below, cycles 8 and 9 experienced higher background levels than the previous two cycles. For this reason, we elected to analyze them separately initially to ensure our results were consistent with previous findings. A future work will combine the entire 3.5\,Ms of all four \nus cycles data to achieve the deepest hard X-ray extragalactic survey to date and enable hard X-ray variability studies covering a five-year baseline.\\
\indent This paper is structured as follows: Section \ref{sec:data,2} describes the \nus data reduction. Section \ref{sec:sims} discusses the \nus simulations, including the reliability and completeness of our results. Section \ref{sec:nus_catalog} details the \nus source catalog. Section \ref{sec:xmm_sec5} describes the \xmm data reduction, source detection, and sensitivity. Section \ref{sec:mw_sec6} discusses the process of finding multiwavelength counterparts, including the X-ray vs optical properties and redshifts. Finally, Section \ref{sec:disc_sec7} discusses the \logn and the CT-fraction, while Section \ref{sec:conc_sec8} lays out our conclusions. \\
\indent All uncertainties listed in the paper are at a 90\% confidence level unless otherwise stated. All magnitudes listed are AB magnitudes. This work adopts the standard cosmological parameters: $\langle H_0\rangle$ = 70\,km\,s$^{-1}$\,Mpc$^{-1}$, $\langle\Omega_M\rangle$ = 0.3, and $\langle\Omega_{\Lambda}\rangle$ = 0.7.

\section{\textit{NuSTAR} Data Processing} \label{sec:data,2}
\nus observed the \jwst NEP TDF 23 times during \nus GO cycles 8 and 9 (PI: Civano, ID: 8180 and ID: 9267, respectively), for a total of $\sim$2.0\,Ms. Additionally, \xmm provided a quasi-simultaneous observation for each of the seven \nus epochs during cycles 8 and 9, totaling 166\,ks. These data are in addition to the $\sim$1.6\,Ms and 62\,ks of \nus and \xmm time, respectively, during \nus cycles 5 and 6 (PI: Civano, ID: 5192 and ID: 6218; 2 yr program). As time-domain science is the main goal, each \nus epoch was taken approximately 3 months apart, amounting to 18 months between the first observation of cycle 8 and the last of cycle 9. The details of each observation can be found in Table \ref{tab:obs_info}. We reduced the \nus data from cycles 8 and 9 using the same approach as was used for cycles 5 and 6. This method is summarized below and additional information can be found in \citetalias{Zhao2021} and \citetalias{Zhao2024}. \\
\indent As mentioned in \citetalias{Zhao2024}, the XMM exposure from Cycle 6 (ObsID: 0870860301) was lost in its entirety due to extreme background levels. Because of this, the \nus and \xmm teams granted our program another round of observations taken in August of 2022 (the first observations listed in Table \ref{tab:obs_info}). By that point, the analysis of \citetalias{Zhao2024} was complete and these observations were left to be included in this work. Therefore, this work includes 13 observations considered as cycle 8 and 13 from cycle 9, for a total of 26 observations.

\subsection{Data Reduction} \label{sec:data_red}
The \nus data were reduced using HEASoft v6.32, \nus Data Analysis Software (NuSTARDAS) v2.1.2, and CALDB v20230718. The \texttt{nupipeline} script was used to calibrate, clean, and screen the level 1 raw data. \\
\indent Following the procedure of \cite{Civano2015}, \citetalias{Zhao2021}, and \citetalias{Zhao2024}, we removed the time periods where the count rate from the 3.5$-$9.5\,keV full-field light curves was at least two times higher than the average count rate. This band has been used in many \nus surveys because it is where radiation from solar flares is the most prevalent. \\
\indent Due to extreme solar activity taking place during the cycle 8 observations, an unusually high amount of time was removed compared to the other three cycles. Cycle 8 had $\sim$13\% of the total observation time removed, while cycle 9 only had $\sim$4\%. Once the flares were removed, we re-ran the \texttt{nupipeline} script on the good time intervals (GTI). Since \nus has bright instrumental emission lines between 24 and 35\,keV, this band was avoided in our survey. Therefore, we proceeded with our analysis using these five energy bands: 3$-$8\,keV, 3$-$24\,keV, 8$-$16\,keV, 8$-$24\,keV, and 16$-$24\,keV.

\renewcommand*{\arraystretch}{1.2}
\begin{table*}
    \caption{Details of the individual observations of \nus and \xmm from cycles 8 and 9. We report the targeting coordinates and the exposures after cleaned for background flares. }
    \centering
    \label{tab:obs_info}
    \begin{tabular}{ccccc|ccccc}
    \hline \hline
    \textbf{ObsId} & \textbf{Date} & \textbf{RA} & \textbf{DEC} & \textbf{Exp.} &  \textbf{ObsId} & \textbf{Date} & \textbf{RA} & \textbf{DEC} & \textbf{Exp.} \\
       &   &  (deg) & (deg) & (ks) & &   &  (deg) & (deg) & (ks) \\
    \hline
    Cycle 8 \nus & & & & & \xmm \\
    \hline
    60666013002 & 2022-08-27 & 260.4542 & 65.7231 & 74.6  & 0870860501 & 2022-08-27 & 260.6917 & 65.8711 & 21.0 \\
    60666014002 & 2022-08-29 & 260.6625 & 65.8422 & 43.3 \\
    60666015002 & 2022-09-02 & 260.9833 & 65.7425 & 64.8 \\
    \hline
    60810001002 & 2022-11-26 & 260.9000 & 66.0314 & 113.8 & 0913590101 & 2022-11-27 & 260.7208 & 65.7700 & 13.0 \\
    60810002002 & 2022-11-29 & 260.3583 & 65.9953 & 113.7 \\
    60810003002 & 2022-12-02 & 260.7083 & 65.8083 & 112.9 \\
    \hline
    60810004002 & 2023-02-24 & 260.7542 & 65.7928 & 119.1 & 0913590501 & 2023-02-25 & 260.7208 & 65.7700 & 21.8 \\
    60810005002 & 2023-02-26 & 261.1167 & 65.8211 & 117.0 \\
    60810006002 & 2023-02-28 & 260.3375 & 65.7958 & 106.3 \\
    \hline
    60810007002 & 2023-05-21 & 260.4417 & 65.7450 & 73.8 \\
    60810008002 & 2023-05-31 & 261.0542 & 65.7481 & 73.1 & 0913590601 & 2023-05-31 & 260.7500 & 65.8500 & 31.1 \\
    60810009002 & 2023-05-23 & 260.7792 & 65.8186 & 23.5 \\
    60810009004 & 2023-05-26 & 260.7708 & 65.8208 & 47.6 \\
    \hline
    \textbf{Total} & & & & 1083.5 & \textbf{Total} & & & & 86.9 \\
    \hline
    \hline 
    Cycle 9 \nus & & & & & \xmm \\
    \hline
    60910001002 & 2023-08-15 & 261.0792 & 66.0464 & 45.7 \\
    60910001004 & 2023-08-30 & 261.0958 & 66.0478 & 51.6 \\
    60910002002 & 2023-08-19 & 260.6750 & 66.0481 & 77.4 \\
    60910003002 & 2023-08-23 & 260.2500 & 66.0478 & 74.1 & 0931420701 & 2023-08-23 & 260.6875 & 65.8183 & 19.1 \\
    \hline
    60910004002 & 2023-11-10 & 261.0833 & 65.8031 & 80.1 \\
    60910005002 & 2023-11-14 & 260.6625 & 65.8047 & 83.4 & 0931420101 & 2023-11-15 & 260.6875 & 65.8183 & 15.5 \\
    60910006002 & 2023-11-21 & 260.2583 & 65.8131 & 85.4 \\
    \hline
    60910007002 & 2024-02-16 & 261.1458 & 65.9183 & 77.4 \\
    60910008002 & 2024-02-19 & 260.7167 & 65.9158 & 76.9 & 0931420501 & 2024-02-21 & 260.7208 & 65.7700 & 31.5 \\
    60910009002 & 2024-02-18 & 260.3083 & 65.9117 & 77.6 \\
    \hline 
    60910010002 & 2024-04-30 & 260.3417 & 65.7439 & 68.0 \\
    60910011002 & 2024-05-02 & 261.1542 & 65.7433 & 67.0 & 0931420601 & 2024-05-02 & 260.7500 & 65.8500 & 13.0 \\
    60910012002 & 2024-05-06 & 260.7917 & 65.7403 & 66.9\\
    \hline
    \textbf{Total} & & & & 931.5 & \textbf{Total} & & & & 79.1 \\
    \hline
    \hline 
    \end{tabular}
\end{table*}

\subsection{Exposure Map} \label{sec:exp_maps}
\begin{figure*}
    \centering
    \includegraphics[width=0.5\linewidth]{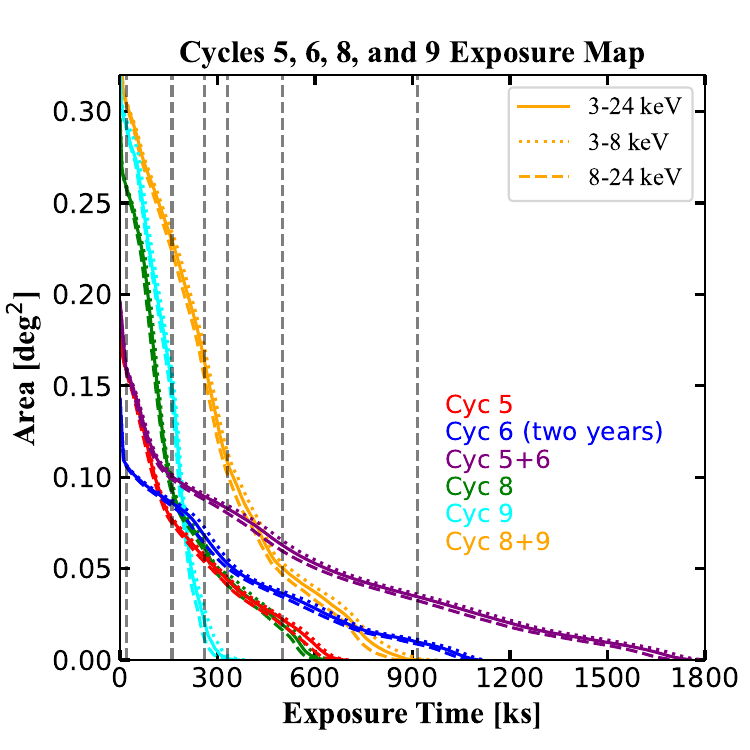}
    \caption{The cumulative area covered by the \nus NEP TDF survey as a function of the FPMA+FPMB vignetting corrected exposure times. The cycle 5 curve is in red, cycle 6 (two year program) in blue, the combined cycles 5 and 6 in purple, cycle 8 in green, cycle 9 in cyan, and  combined cycles 8 and 9 in orange. The 3$-$24\,keV curve is plotted as a solid line, the \thei as a dotted line, and the \eitw as a dashed line. The grey vertical lines enclose the five exposure bins used to determine the reliability of the cycles 8+9 survey (see Section \ref{sec:rel_and_compl}).}
    \label{fig:exp_map}
\end{figure*}

Using the NuSTARDAS tool \texttt{nuexpomap}, we created the vignetting corrected exposure map for the energy bands 3$-$8\,keV, 3$-$24\,keV, and 8$-$24\,keV. The 8$-$24\,keV band exposure map was also used for the 8$-$16\,keV and 16$-$24\,keV bands, as only marginal differences exist between the three maps. The cycle 8 and 9 exposure maps were created by summing all individual observations from the respective cycle into one mosaic. Moreover, the two focal plane modules, FPMA + FPMB, were combined to maximize the sensitivity of the survey. Figure \ref{fig:exp_map} displays the exposure map curves for cycles 8 and 9 individually, as well as the combined exposure map for all 26 observations from cycles 8 and 9. This figure also makes clear the different observing strategies for cycles 5 and 6 versus cycles 8 and 9. The former focused on a deep survey ($\sim$1.6\,Ms) covering a smaller area ($\sim$0.16 deg$^2$).  Instead, cycles 8+9 have a combined exposure of $\sim$900\,ks at their deepest, but cover an area approximately double that of cycles 5+6 ($\sim$0.31 deg$^2$).

\subsection{Mosaic Creation}
\begin{figure}
    \hspace{-1.5cm}
    %\centering
    \includegraphics[width=1.4\linewidth]{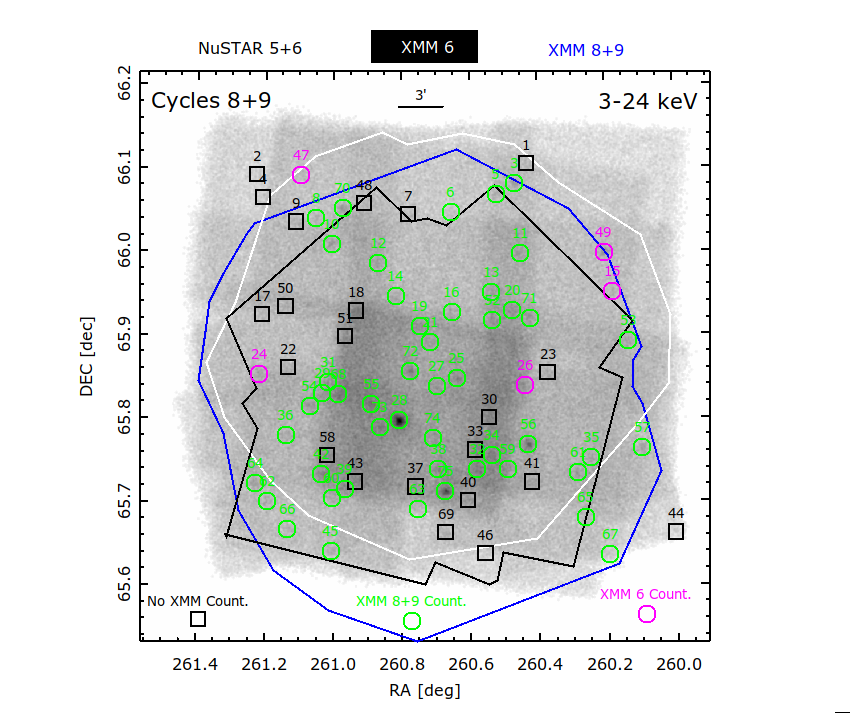}
    \caption{The combined \nus cycles 8+9 mosaic in the \thtw band. All 75 detected sources are labeled. The green circles (25$\arcsec$ radius) represent the 48 sources with \xmm counterparts from cycles 8+9, the magenta circles (25$\arcsec$ radius) represent the 5 sources with \xmm counterparts from cycle 6, and the black squares (45$\arcsec$ width) represent the 22 \nus sources without \xmm counterparts. See Section \ref{sec:xmm_nus_crossmatch} for more details. Also plotted are footprints from the \nus cycles 5+6 survey (black), the \xmm cycle 6 survey (white), and the \xmm cycles 8+9 survey (blue).}
    \label{fig:cyc89_mosaic}
\end{figure}

By combining together the 13 and 13 observations from cycle 8 and 9, respectively, we created mosaics for source detection in five different energy bands: 3$-$8\,keV, 3$-$24\,keV, 8$-$16\,keV, 8$-$24\,keV, and 16$-$24\,keV. The \texttt{Xselect} tool was used to filter each observation (summed FPMA+FPMB) into these five energy bands. These individual observations were then summed into a single mosaic using the \texttt{Ximage} tool. The cycle 8 and 9 mosaics were summed to reach the maximum sensitivity at this increased area (see Figure \ref{fig:exp_map}). The \nus COSMOS survey determined that the typical \nus astrometric offset is on the order of 1$-$7$\arcsec$ \citep{Civano2015}. However, since there is only one bright source in the field of view (FoV) that could be used for the astrometric correction, we could not perform astrometric correction when combining the observations. Figure \ref{fig:cyc89_mosaic} shows the combined mosaic of all 26 observations.

\subsection{Background Map} \label{sec:bkg_maps}
Background maps were used for source detection on both the real data and the simulations. The \nus background is spatially non-uniform across the FoV and can vary across different observations. Therefore, the background from each observation must be modeled separately to reproduce the observed background as accurately as possible. As was done by \citetalias{Zhao2021}, \citetalias{Zhao2024}, and many \nus extragalactic surveys before, we created the background maps using \texttt{nuskybgd}\footnote{\url{https://github.com/NuSTAR/nuskybgd}} \citep{Wik2014}. We merged the background maps for every observation in cycles 8 and 9 to produce a mosaic map for each cycle. Then, we made a combined cycles 8 and 9 background mosaic using all 26 observations. \\
\indent In order to verify our background maps are accurate, we compared the number of counts in the observations with those in the corresponding maps. Considering the background dominates over the sources' flux for the vast majority of the FoV, excluding the location of the few bright sources, the observed number of counts in the observations should be nearly the same as the background maps. To test this, we evenly divided the FoV into 64 circular regions with a radius of 45$\arcsec$ and extracted the counts from both the observations and background maps. We then calculated the percent difference between the two images using: (Data - Bkg) / Bkg. Based on the results from \citetalias{Zhao2021}, \citetalias{Zhao2024}, and the \nus COSMOS survey \citep{Civano2015}, we expect a mean difference around $\approx$0$-$2\% and a standard deviation $\approx$12$-$14\% for each detector. Our cycle 9 background maps performed well, with a mean of 0.1\% and 2.3\%, and standard deviations of 12.9\% and 12.5\% for detectors FPMA and FPMB, respectively, in the \thtw band. However, as stated in Section \ref{sec:data_red}, our cycle 8 observations were affected by unusually high amounts of solar activity. \texttt{nuskybgd} was able to account for most of this, but not all. Consequently, detectors FPMA and FPMB yielded a mean percent difference of 4.6\% and 6.1\%, and standard deviations of 15.7\% and 16.3\%. These larger differences were accounted for in the determination of the reliability thresholds used for source detections (see Section \ref{sec:rel_and_compl}). \\

\section{\nus Simulations} \label{sec:sims}
We performed 1200 comprehensive simulations for each \nus observation from cycle 8, cycle 9, and the combined cycles 8 and 9, covering five energy bands: 3$-$8\,keV, 3$-$24\,keV, 8$-$16\,keV, 8$-$24\,keV, and 16$-$24\,keV. These detailed simulations allowed us to 1) determine the reliability and completeness of the source detection, 2) calculate the sensitivity, and 3) determine the robustness of our source detection technique by comparing the source properties, such as flux and position.

\subsection{Generating Simulated Data} \label{sec:sims_generate}
We followed the same approach used in \citetalias{Zhao2021} and \citetalias{Zhao2024} to create the simulated data. To briefly summarize, these simulations were created by generating mock sources and randomly placing them on the background maps described in Section \ref{sec:bkg_maps}. The fluxes of these sources were randomly assigned based on the Log$N$-Log$S$ distribution measured in \cite{Treister2009}. The flux limit input into the simulations was about 10$\times$ fainter than that of the expected flux limit of the three surveys. In the 3-24\,keV band, the applied flux limits for the cycles 8, 9, and 8+9 simulations were 3 $\times$ 10$^{-15}$ erg cm$^{-2}$ s$^{-1}$, 4 $\times$ 10$^{-15}$ erg cm$^{-2}$ s$^{-1}$, and 2.5 $\times$ 10$^{-15}$ erg cm$^{-2}$ s$^{-1}$, respectively. These limits were chosen to avoid a large number of false matches. The flux for each band (3$-$8\,keV, 8$-$16\,keV, 8$-$24\,keV, and 16$-$24\,keV) was extrapolated from the \thtw flux using a power-law model with $\Gamma$ = 1.80 and Galactic absorption across the NEP TDF of N$_{\rm H}$ = 3.4 $\times$ 10$^{20}$ cm$^{-2}$ \citep{HI4PI_collab2016}. Then, these fluxes were converted into count rates using the count-rate-to-flux conversion factors (CF) calculated with WebPIMMS\footnote{\url{https://heasarc.gsfc.nasa.gov/cgi-bin/Tools/w3pimms/w3pimms.pl}}, implementing the model above. The conversion factors used for the 3$-$8\,keV, 3$-$24\,keV, 8$-$16\,keV, 8$-$24\,keV, and 16$-$24\,keV bands were 3.39, 4.86, 5.17, 7.08, and 16.2 $\times$ 10$^{-11}$ erg cm$^{-2}$ counts$^{-1}$, respectively. Once the simulations were completed, the observations from each exposure were summed for both FPMA and FPMB in all five energy bands. These mosaics were then merged into the final FPMA+B simulated mosaics for each observation.

\subsection{Source Detection on Simulated Data} \label{sec:sim_sou_det}
Source detection was carried out on the cycles 8+9 simulated FPMA+B mosaics following the procedure established in \cite{Mullaney2015} and used in \citetalias{Zhao2021} and \citetalias{Zhao2024}. A brief summary of the procedure goes as follows. First, \texttt{SExtractor} \citep{Bertin+Arnouts1996} was used to perform source detection on the false-probability maps. These maps measure the probability (P$_{false}$) that any signal is caused by fluctuations of the background, rather than a true source. To create these probability maps, we used the incomplete Gamma function to compare the smoothed simulated maps and the real background mosaics for each pixel. We set the detection limit to P$_{false} \leq$ 10$^{-2.5}$ (i.e., $\sim$3\,$\sigma$). Then, the Poisson probability (P$_{random}$) was calculated at the position of each detected source and used to indicate if the detection was caused by a random background fluctuation. This probability was calculated by extracting the total and background counts from the simulated and background maps using a 20$\arcsec$ extraction radius. Then, we defined the maximum likelihood (DET\_ML) of every detection by inverting the logarithm of the Poisson probability. Therefore, DET\_ML = -lnP$_{random}$. This means a small value of P$_{random}$ corresponds to a high DET\_ML, and therefore the detection is unlikely to be caused by a random background fluctuation. \\
\indent The measured source counts may be affected by nearby sources less than 90$\arcsec$ away due to the 85-90\% encircled energy fraction (EEF) of the \nus point spread function (PSF). To account for this, we implemented a deblending process on the detected sources, following the procedure in \cite{Mullaney2015}. These deblended source counts and background counts were then used to calculate updated DET\_ML values. The updated DET\_ML for every simulated detection was matched with the sources from the input catalog using a search radius of 30$\arcsec$. The top two lines of Table \ref{tab:rel_thresh} list the average number of sources detected and matched to the input catalogs for all five energy bands. 

\subsection{Reliability and Completeness} \label{sec:rel_and_compl}
We describe the methods to calculate the reliability and completeness of our simulations, which are used to measure the accuracy and efficiency of the source detection performed. First, reliability is defined as the number of sources matched to the input catalog divided by the total number of sources detected (both must be above a determined DET\_ML threshold):

\begin{equation}
   Rel(DET\_ML\_thresh) = \frac{N_{matched}(\geq DET\_ML\_thresh)}{N_{detected}(\geq DET\_ML\_thresh)}
\end{equation}

In simple terms, if 100 sources with \detml $\geq$ 10 are detected and 90 are matched to the input catalog, then the reliability for the survey is 90\% at \detml = 10. \\
\indent Second, completeness is defined as the number of detected sources matched to the input catalog and above a certain threshold divided by the total number of input sources at a specific flux:

\begin{equation}
   Completeness (flux) = \frac{N_{matched \& \geq Rel\_thresh}(flux)}{N_{input}(flux)}
\end{equation}

\begin{figure*}
    \centering
    \includegraphics[width=0.45\linewidth]{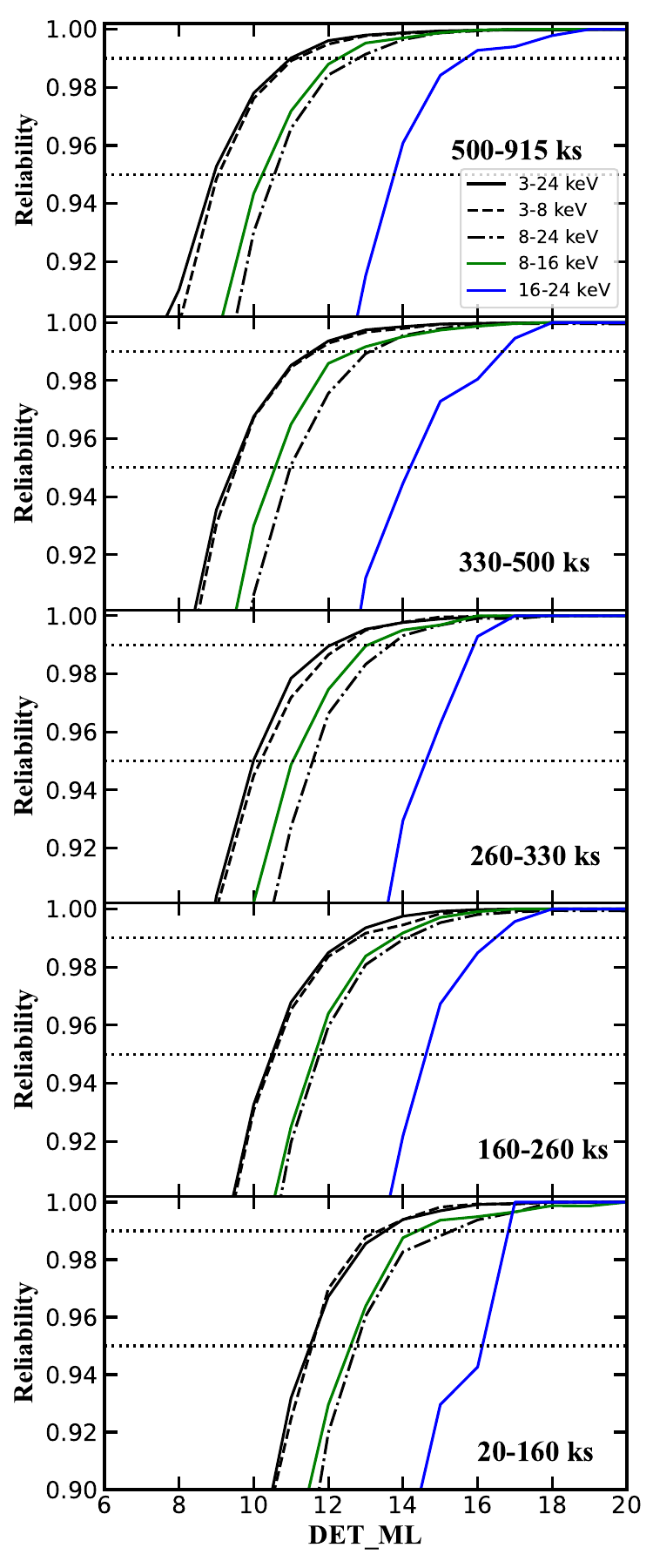} 
    \includegraphics[width=0.45\linewidth]{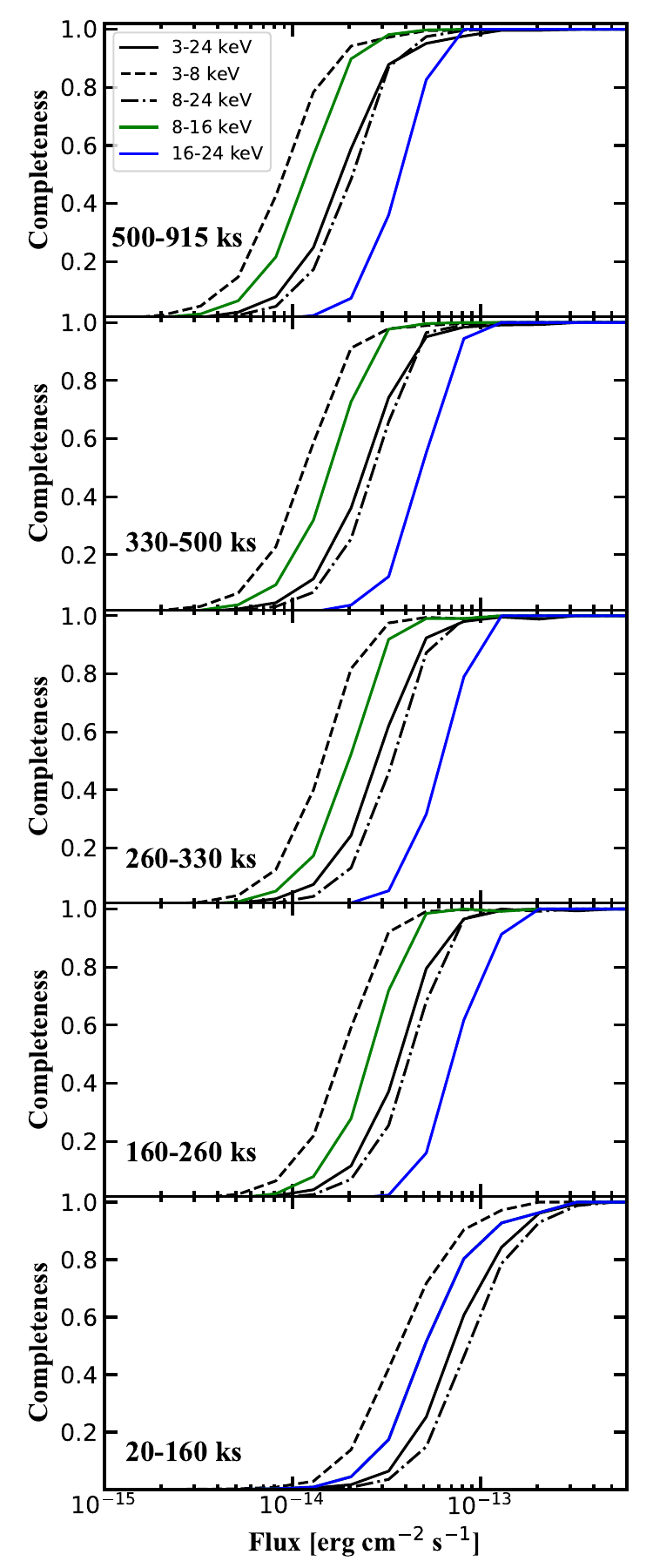}
    \caption{The five plots display each of the five different time bins selected to measure the reliability and completeness of our cycle 8+9 simulations. These bins were selected to provide an equal amount of sources per bin, thus ensuring significant statistics could be measured for every bin. The left plots show the reliability as a function of DET\_ML and the right plots show the completeness at a 99\% reliability level as a function of flux. In all plots, the solid black line represents the \thtw band, the dashed black the 3$-$8\,keV, the dash-dot black line the 8$-$24\,keV, the solid green the 8$-$16\,keV, and the solid blue line the 16$-$24\,keV. On the left, the 95\% and 99\% reliability levels are indicated with horizontal black dotted lines. Note: It is not correct to directly compare the curves from each different energy band. A conversion factor, dependent on the two bands in question, must be applied in order to do so.}
    \label{fig:rel+compl}
\end{figure*}

As an example, if 75 out of the input 100 sources with fluxes greater than 1 $\times$ 10$^{-13}$ erg cm$^{-2}$ s$^{-1}$ are detected above a 95\% reliability threshold, then the completeness at that chosen flux is 75\%. \\
\indent In order to reach a higher reliability level, \detml must increase, which in turn will decrease the completeness of the survey. In this work, we follow the procedure of \citetalias{Zhao2021} and \citetalias{Zhao2024} and select the 95\% and 99\% reliability thresholds. The left panels of Figure \ref{fig:rel+compl} show the cycle 8+9 reliability as a function of \detml for all five energy bands. The right column shows the completeness versus the flux for the same five bands and exposure bins. The completeness is calculated for sources above a 99\% reliability threshold, thus implying a $\sim$1\% spurious detection rate. \\
\indent It is known that reliability and completeness are largely dependent on the effective exposure time. Considering how our observing strategy covers the NEP TDF non-uniformly (see Figure \ref{fig:cyc89_mosaic}), assigning one reliability and completeness curve to the entire FoV would produce misleading results. Therefore, we followed \citetalias{Zhao2021} and \citetalias{Zhao2024} and split all the detections into five bins based on exposure time, ensuring that every bin had an equal number of detections. For the cycle 8+9 survey simulations, these exposure bins are as follows: 20$-$160\,ks, 160$-$260\,ks, 260$-$330\,ks, 330$-$500\,ks, and 500$-$915\,ks. This way, each bin had $\sim$26k detections, ensuring a similar level of statistical significance could be achieved for every exposure bin. Additionally, we imposed a lower-limit exposure cutoff at 20\,ks to remove any potential spurious detections located on the edges of the mosaic with minimal exposure time. \\
\indent Figure \ref{fig:rel+compl} shows how the reliability and completeness curves vary for each exposure bin. These differences occurs because the \detml required to reach a given reliability threshold decreases as exposure increases. Moreover, the completeness curves follow a similar trend of moving to the left as the exposure increases because fainter sources become easier to detect in larger numbers with deeper exposures. Table \ref{tab:rel_thresh} reports the 95\% ad 99\% \detml thresholds for every exposure bin, as well as the average number of sources detected and matched for the cycles 8+9 simulations.

\begin{table*}
    \centering
    \caption{The results of the cycles 8+9 simulations. The first two lines show the average number of sources detected and matched to the input catalog (within 30$\arcsec$) per simulation. Lines 3-7 and 8-12 show the \detml thresholds in each exposure bin for a 99\% and 95\% reliability cutoff, respectively. Lines 13-14 show the average number of sources detected above the 99\% and 95\% thresholds per simulation. Lines 15-16 show the number of unique detections above the 99\% and 95\% thresholds for the real cycle 8+9 data.}
    \begin{tabular}{ccccccc}
    \hline
    \hline
    & & & Cycle 8+9  \\
    & \thei & \thtw & \eisi & \eitw & \sitw & \\
    Detections in Simulated Maps & 143 & 148 & 133 & 136 & 117 & \\
    Matched to Input & 98 & 105 & 79 & 79 & 44 & \\
    \hline
    DET\_ML(99\%, 20$-$160\,ks) threshold & 13.4 & 13.5 & 14.4 & 15.3 & 23.4 \\ 
    DET\_ML(99\%, 160$-$260\,ks) threshold & 12.8 & 12.6 & 13.8 & 14.1 & 16.5 \\
    DET\_ML(99\%, 260$-$330\,ks) threshold & 12.4 & 12.1 & 13.1 & 13.7 & 15.9 \\
    DET\_ML(99\%, 330$-$500\,ks) threshold & 11.7 & 11.6 & 12.7 & 13.1 & 16.7 \\
    DET\_ML(99\%, 500$-$915\,ks) threshold & 11.1 & 11.0 & 12.3 & 12.8 & 15.7 \\
    \hline
    DET\_ML(95\%, 20$-$160\,ks) threshold & 11.6 & 11.5 & 12.6 & 12.7 & 17.2 \\
    DET\_ML(95\%, 160$-$260\,ks) threshold & 10.6 & 10.5 & 11.6 & 11.8 & 14.6 \\
    DET\_ML(95\%, 260$-$330\,ks) threshold & 10.2 & 10.0 & 11.1 & 11.6 & 14.6 \\
    DET\_ML(95\%, 330$-$500\,ks) threshold & 9.5 & 9.5 & 10.6 & 11.0 & 14.2 \\
    DET\_ML(95\%, 500$-$915\,ks) threshold & 9.1 & 8.9 & 10.2 & 10.6 & 13.8 \\
    \hline 
    Simulated Maps & & & & & & \\
    N$_{\rm src}$ (DET\_ML$>$99\% reliability threshold) & 29.6 & 36.9 & 14.7 & 13.4 & 1.3 \\
    N$_{\rm src}$ (DET\_ML$>$95\% reliability threshold) & 43.6 & 53.7 & 21.6 & 20.3 & 1.8 \\
    \hline
    Real Data & & & & & & Total \\
    N$_{\rm src}$ (DET\_ML$>$99\% reliability threshold) & 48 & 69 & 18 & 22 & 3 & 75 \\
    N$_{\rm src}$ (DET\_ML$>$95\% reliability threshold) & 80 & 101 & 36 & 38 & 3 & 128 \\
    \hline
    \hline
    \end{tabular}
    \label{tab:rel_thresh}
\end{table*}

\subsection{Sensitivity Curves}
\begin{figure*}
    \centering
    \includegraphics[width=0.5\linewidth]{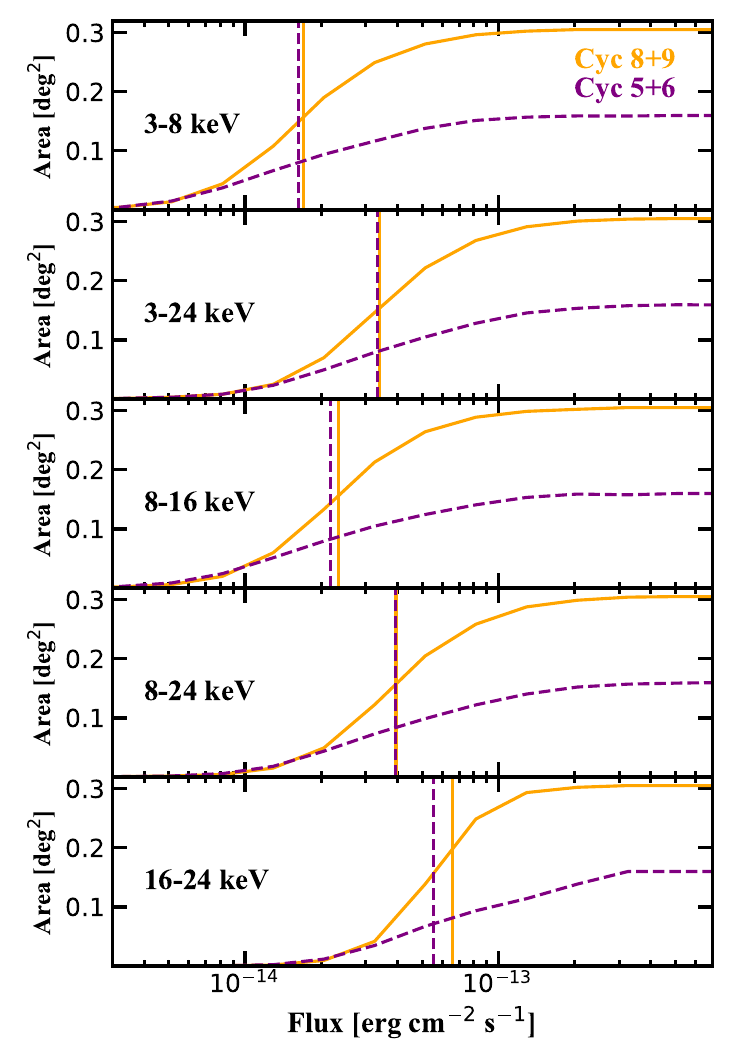}
    \caption{The sky coverage as a function of flux at a 99\% reliability level for all five energy bands. The orange solid lines represent the cycles 8+9 curves while the purple dashed lines represent the cycles 5+6 curves. The vertical lines illustrate the half-area fluxes.}
    \label{fig:area_coverage}
\end{figure*}

\begin{figure*}
    \centering
    \includegraphics[width=0.5\linewidth]{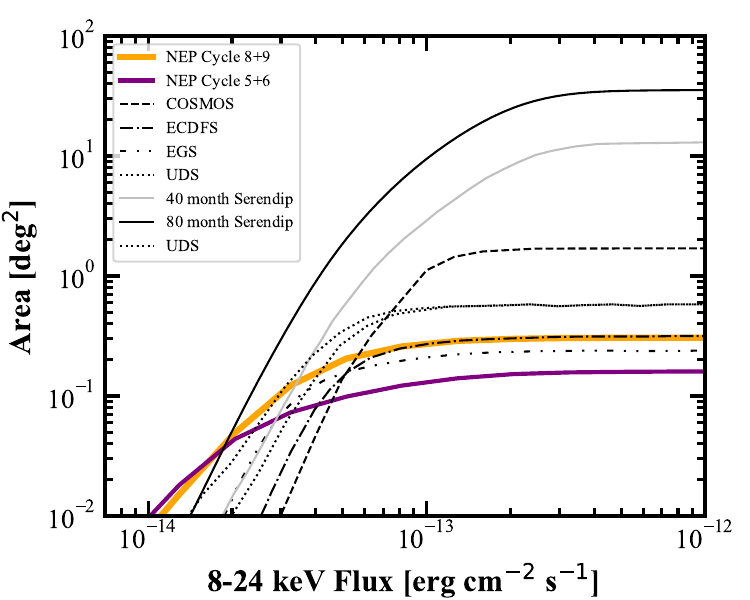}
    \caption{The \eitw sensitivity curves of the NEP TDF survey for cycles 8+9 (orange) and 5+6 (purple). Other \nus surveys included for comparison are the: COSMOS \citep[dashed line;][]{Civano2015}, ECDFS \citep[dash-dotted line;][]{Mullaney2015}, EGS (dash dot dotted line; Aird et al. in preparation), UDS \citep[dotted line;][]{Masini2018a}, 40-month Serendipitous \citep[gray solid line;][]{Lansbury2017}, and the 80-month Serendipitous \citep[black solid line;][]{Greenwell2024}.}
    \label{fig:sens_8-24}
\end{figure*}

The completeness derived above can be used to calculate the sky coverage (or sensitivity) of the survey. The sensitivity at a given flux is the completeness at that flux multiplied by the maximum area covered indicated by the exposure map. Using the example above, if the completeness at a given flux is 75\% and the survey covers 0.1\,deg$^2$, then the survey is sensitive to sources at that flux in only 0.075\,deg$^2$ of the full survey. \\
\indent As stated above (and shown in Figure \ref{fig:rel+compl}), completeness is dependent on the effective exposure. Therefore, the total sensitivity is as well. To account for this, we summed the sensitivity curves for every exposure bin. The areas covered by the different exposure bins are as follows: 0.081 deg$^2$ (20$-$160\,ks), 0.070 deg$^2$ (160$-$260\,ks), 0.053 deg$^2$ (260$-$330\,ks), 0.056 deg$^2$ (330$-$500\,ks), and 0.047 deg$^2$ (500$-$915\,ks). Figure \ref{fig:area_coverage} displays the total sky coverage for all five bands for both cycles 5+6 (purple) and 8+9 (orange). %This plot further emphasizes the different strategies of the two surveys. In particular, the Cycles 8+9 curve covers more area at brighter fluxes (right of the half-area flux vertical line) while the Cycles 5+6 survey is deeper, and thus covers more area in the faint flux regime (left of the half area flux). 
The values for the half-area and 20\%-area fluxes for both surveys can be found in Table \ref{tab:half_area_flux}. \\
\indent As \nus is the only instrument capable of observing the \eitw band with high sensitivity, we compare the fluxes reached in this survey with those of previous \nus surveys in Figure \ref{fig:sens_8-24}. The cycles 5+6 survey from \citetalias{Zhao2024} remains the deepest contiguous \nus survey, with the cycles 8+9 survey reaching the second lowest fluxes while covering a larger area. Also shown are the \nus COSMOS \citep{Civano2015}, ECDFS \citep{Mullaney2015}, EGS, 40-month Serendipitous \citep{Lansbury2017}, UDS \citep{Masini2018a}, and the 80-month Serendipitous \citep{Greenwell2024}.

\begin{table}[]
    \centering
    \begin{tabular}{ccc}
        \hline \hline
        Energy & Half-area & 20\%-area \\
        \hline
         \nus & \\
         \hline
         keV & 10$^{-14}$ erg cm$^{-2}$ s$^{-1}$ & 10$^{-14}$ erg cm$^{-2}$ s$^{-1}$ \\
         \hline
          & Cycles 5+6 / 8+9 & Cycles 5+6 / 8+9 \\
          \hline 
          3$-$8 & 1.7 / 1.7 & 0.7 / 0.9 \\
          3$-$24 & 3.3 / 3.4 & 1.6 / 1.9 \\
          8$-$16 & 2.1 / 2.3 & 1.0 / 1.3 \\
          8$-$24 & 3.8 / 4.0 & 1.7 / 2.2 \\
          16$-$24 & 6.6 / 7.1 & 3.1 / 3.6 \\
          \hline \hline
          \xmm \\
          \hline 
          keV & 10$^{-15}$ erg cm$^{-2}$ s$^{-1}$ & 10$^{-15}$ erg cm$^{-2}$ s$^{-1}$ \\
          \hline
          & Cycle 6 / Cycles 8+9 & Cycle 6 / Cycles 8+9 \\
          \hline
          0.5$-$2 & 0.9 / 0.9 & 0.6 / 0.7 \\
          2$-$10 & 6.3 / 6.9 & 4.0 / 4.2 \\
          \hline \hline
         
    \end{tabular}
    \caption{The half-area and 20\%-area fluxes listed for the cycles 5+6 and 8+9 surveys.}
    \label{tab:half_area_flux}
\end{table}

\subsection{Fluxes} \label{sec:nus_fluxes}
\begin{figure}
    \centering
    \includegraphics[scale=0.65]{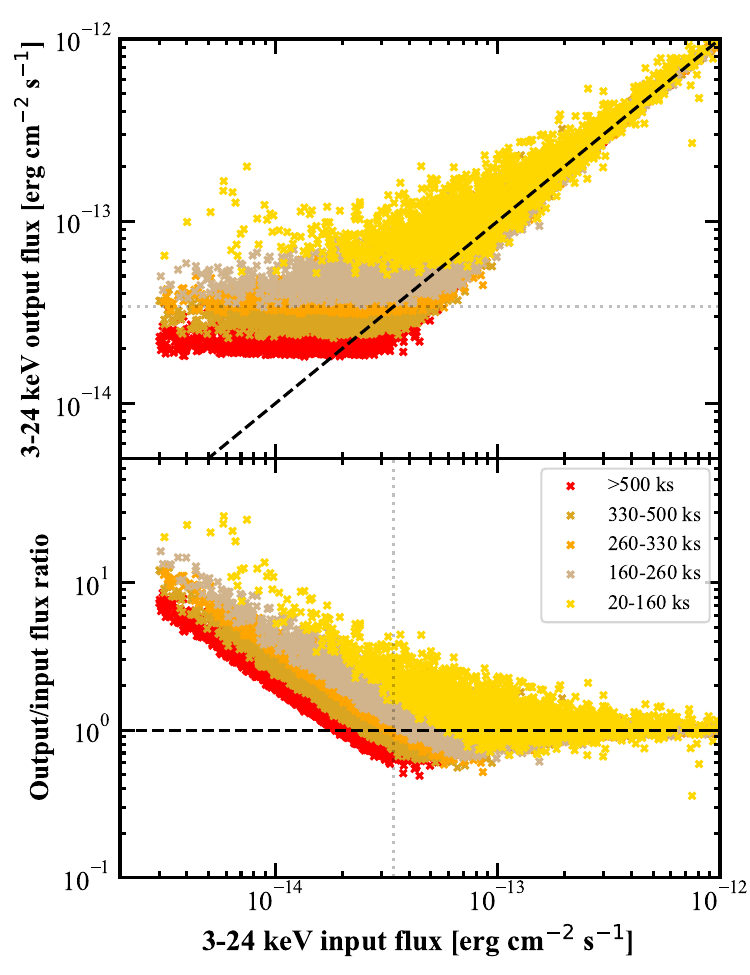}
    \caption{\textbf{Top:} The 3-24\,keV measured fluxes plotted against the input fluxes. Only the simulated sources detected above the 99\% reliability threshold are displayed. The red sources have a simulated exposure $>$500\,ks, gold represents sources with 330$-$500\,ks, orange for 260$-$330\,ks, tan for 160$-$260\,ks, and yellow for 20$-$160\,ks. \textbf{Bottom:} The ratio between the output and input fluxes with the colors representing the same exposures as above. The excess shown at lower fluxes can be ascribed to Eddington bias.
    In both plots, the grey dotted lines represent the 50\%-area flux from this survey in the \thtw band.}
    \label{fig:sims_flux}
\end{figure}

In order to establish the accuracy of the flux measurements, we compared the output versus input fluxes for all detections in every band. The CIAO tool \texttt{dmextract} \citep{Fruscione2006} was used to extract the source counts and deblended background counts for every matched source. The effective exposure for each source was measured using the exposure maps discussed in Section \ref{sec:exp_maps}. We converted the detected net counts into fluxes using the exposure time at each source position and the count-rate-to-flux CFs listed in Section \ref{sec:sims_generate}. Each count extraction used a 20$\arcsec$ region. Therefore, to convert from this flux to the total flux, we used a factor of F$_{20\arcsec}$ / F$_{tot}$ = 0.32, which was derived from the \nus PSF\footnote{\url{https://heasarc.gsfc.nasa.gov/docs/nustar/NuSTAR_observatory_guide-v1.0.pdf}}. \\
\indent Figure \ref{fig:sims_flux} displays the 3-24\,keV input fluxes compared with the measured output fluxes of the sources detected above the 99\% reliability threshold from the cycles 8+9 simulations. The discrepancy at lower fluxes is caused by the Eddington bias, i.e., the only faint sources that can be detected are those with positive noise fluctuations. This excess is related to the detection limits of this survey.

\subsection{Positional Uncertainty}
\begin{figure}
    \centering
    \includegraphics[scale=0.7]{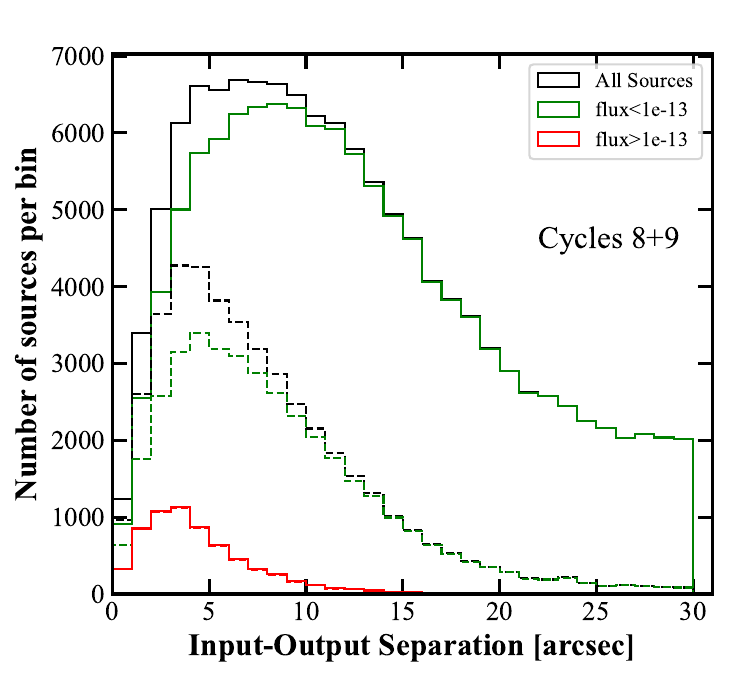}
    \caption{The distributions of the difference between the input and output positions of sources detected in the \thtw band in the cycles 8+9 simulations. The solid lines represent all detected sources while the dashed lines represent only the sources detected above a 99\% confidence threshold. The black lines represent all sources while the red and green lines represent sources with \thtw fluxes above and below 1 $\times$ 10$^{-13}$ erg cm$^{-2}$ s$^{-1}$, respectively.}
    \label{fig:pos_unc}
\end{figure}

Figure \ref{fig:pos_unc} shows the difference in positions for the input and output sources detected in the 3$-$24\,keV band from our cycles 8+9 simulations. These separation histograms were created using a Rayleigh distribution \citep{Pineau2017}. The sample containing all sources (black solid line) had a mean separation of 12.5$\arcsec$, while the sample of only sources detected above the 99\% reliability threshold (black dashed line) had a mean separation of 7.8$\arcsec$. The separation for bright sources ($\sigma_{99\%,bright}$ = 4.5$\arcsec$) was found to be smaller than that for faint sources ($\sigma_{99\%,faint}$ = 8.4$\arcsec$). We note that these numbers are consistent with separations found in previous \nus extragalactic surveys, although slightly higher ($\sim$1-2$\arcsec$ larger). This is likely caused by one of two possibilities (or a combination of the two): 1) the decreased net exposure in this survey compared to the cycles 5+6 survey, or 2) the previously stated high background experienced during the cycles 8 and 9 observations. However, these results are still sufficient to be used as the positional uncertainty for the detections in the real data.

\section{\nus Source Catalog} \label{sec:nus_catalog}
\begin{figure}
    %\centering
    \includegraphics[scale=0.6]{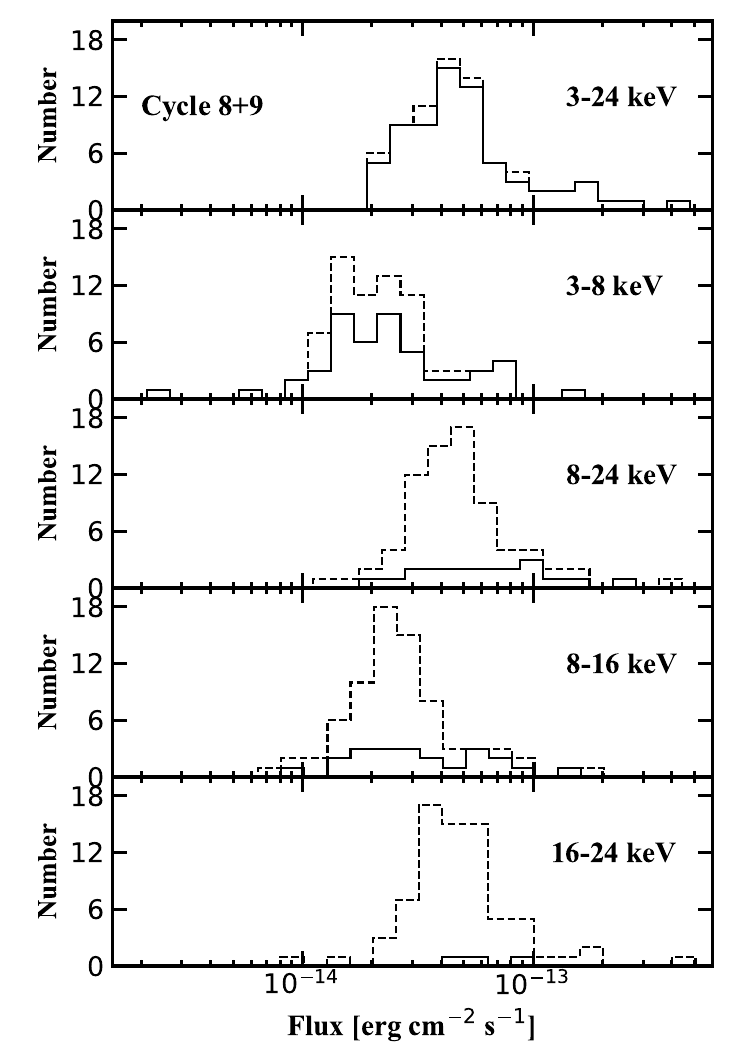}
    \caption{The flux distributions for all five energy bands for the Cycles 8+9 survey. The solid lines represent the sources detected above the 99\% reliability threshold. The dashed line represents a 90\% confidence flux upper limit for all sources not detected and those detected but below the 99\% threshold.}
    \label{fig:flux_distr}
\end{figure}

The source detection was performed in each of the five energy bands on the FPMA+B mosaics in order to maximize the signal-to-noise ratio (SNR). For every source, we measured the coordinates, source counts, background counts, DET\_ML, and vignetting-corrected exposure time. The sources detected above a 99\% reliability threshold in at least one of the five energy bands were recorded into the \nus master catalog. The coordinates listed in these catalogs correspond to the detection in the energy band with the highest DET\_ML. Figure \ref{fig:cyc89_mosaic} displays the positions of the sources detected in the cycles 8+9 survey. \\
\indent Table \ref{tab:rel_thresh} lists the number of detections above the 95\% and 99\% reliability thresholds for the cycles 8+9 survey: 128 sources were detected above a 95\% reliability level and 75 sources above a 99\% threshold. This corresponds to $\sim$6 spurious detections above the 95\% level and $\sim$1 above the 99\% level. For comparison, the cycles 5+6 survey detected 60 and 30 total sources, with 2$-$3 and 1 spurious detection, respectively. Unlike in \citetalias{Zhao2021} and \citetalias{Zhao2024} which primarily used the 95\% reliability threshold, in the remainder of this paper we will focus only on the sources detected above the 99\% reliability threshold due to the high solar background levels. \\
\indent Table \ref{tab:distr_sourc_det} lists the number of sources from cycles 8+9 detected in different energy band combinations. F, S, and H represent sources detected above the 99\% reliability threshold in the full (3$-$24\,keV), soft (3$-$8\,keV), and hard (8$-$16\,keV, 8$-$24\,keV, 16$-$24\,keV) energy bands. Additionally, f, s, and h represent sources detected above the 95\% level but not the 99\% level for the same energy bands. \\ %Out of the 75 sources from this survey, 16 (or 21\%) were detected in the soft, full, and hard band. 15 sources (20\%) were detected in the full and soft band without a detection above the 95\% threshold in the hard band. Meanwhile, only 5 (7\%) sources were detected in the full and hard bands without a 95\% detection in the soft band. \\
\indent The net counts of detections were calculated by subtracting the deblended background counts from the total counts extracted with a 20$\arcsec$ circular region. We note that if a source was not detected in a specific energy band, the extracted background counts were not deblended. The next step converted the net count rates in each energy band into fluxes by dividing them by the vignetting-corrected exposure times and multiplying by the correction factors listed in Section \ref{sec:sims_generate}. Additionally, the fluxes needed to be corrected from the aperture fluxes to the total fluxes using the aperture correction factor, (i.e., $F_{20\arcsec}$ / $F_{tot}$ $\sim$ 0.32). Using Equations (9) and (12) from \cite{Gehrels1986} (with S = 1), we calculated the 1\,$\sigma$ net count rate uncertainties and flux uncertainties for the sources detected above the 99\% level. For sources not detected or detected below the 99\% threshold, a 90\% flux upper limit was determined using Equation (9) and S = 1.645. The flux distributions of these sources for all five energy bands can be viewed in Figure \ref{fig:flux_distr}. \\
\indent Due to the faint fluxes reached by this survey (see Figure \ref{fig:sens_8-24}), there is a potential concern that multiple undetected sources could fall within one extraction radius, thus contaminating the results. However, the LogN$-$LogS plots shown in Section \ref{sec:logn-logs} prove our results are unlikely to be affected by such faint sources. According to the \cite{Ueda2014} 8-24\,keV model, one expects about 400 sources per deg$^2$ with a flux of 1$\times$ 10$^{-14}$ erg cm${-2}$ s$^{-1}$ (the approximate sensitivity of our survey). Therefore, the average distance between each source in an extragalactic field with no structure like the NEP is about 3$\arcmin$. This is much greater than the FWHM of \textit{NuSTAR} (7.5$\arcsec$\footnote{\url{ https://heasarc.gsfc.nasa.gov/docs/nustar/nustar.html}}). Moreover, the encircled energy fraction (EEF) of \textit{NuSTAR} at 100$\arcsec$ is $\sim$80\% \citep{Harrison2013}. Combining these two facts, we do not expect our source detection and flux estimation to be impacted by these unseen sources. \\
\indent The 99\% reliability catalog is made public with the publication of this paper. Table \ref{Table:nus_cat_info} in the Appendix provides the description of every column in the catalog.

\begin{table}
    \centering
    \begin{tabular}{cc}
        \hline \hline
        Energy & Cycle 8+9 \\
        \hline
        F+S+H & 16 (21\%) \\
        F+S+h & 12 (16\%) \\
        F+s+H & 2 (3\%) \\
        f+S+H & 1 (1\%) \\
        %F+s+h & 0 (0\%) \\
        %f+S+h & 0 (0\%) \\
        %f+s+H & 0 (0\%) \\
        F+S & 15 (20\%) \\
        F+H & 5 (7\%) \\
        %S+H & 0 (0\%) \\
        F+s & 9 (12\%) \\
        F+h & 3 (4\%) \\
        f+S & 1 (1\%) \\
        %f+H & 0 (0\%) \\
        S+h & 1 (1\%) \\
        %s+H & 0 (0\%) \\
        F & 7 (9\%) \\
        S & 2 (2\%) \\
        H & 1 (1\%) \\
        \hline
        Total & 75 \\
        \hline \hline
    \end{tabular}
    \caption{Number of the sources detected above the 99\% confidence level in the Cycles 8+9 survey. F(f), S(s), and H(h) represent the full (3$-$24\,keV), soft (3$-$8\,keV), and hard (8$-$16\,keV, 8$-$24\,keV, 16$-$24\,keV) energy bands. Capital letters represent sources detected above the 99\% reliability level while lowercase letters represent sources only detected above the 95\% reliability level.}
    \label{tab:distr_sourc_det}
\end{table}

\section{XMM-Newton NEP-TDF Survey} \label{sec:xmm_sec5}
Following the strategy used for \nus cycle 6, \xmm observations, approximately 20\,ks in duration, were taken simultaneously with each epoch of \nus observations during cycles 8 and 9. These XMM observations allow for a complete broadband spectral analysis of these sources in the 0.3 to 24\,keV energy range. In addition, the superior spatial resolution of XMM compared to \textit{NuSTAR} ($\sim$5$\arcsec$ vs $\sim$30$\arcsec$) allows for more accurate positional determinations. \\
\indent As discussed in Section \ref{sec:bkg_maps}, the solar activity during the cycle 8 and 9 observations was much higher than in cycles 5 and 6. This affected the \xmm data as well as the \textit{NuSTAR} data. In cycle 8, the four XMM observations lost a total of 47\,ks due to flares, resulting in $\sim$40\,ks of net exposure. The cycle 9 observations only lost 21\,ks, thus totaling $\sim$58\,ks of net exposure. The details of each XMM observation are displayed in Table \ref{tab:obs_info}. The cycle 8 XMM observations covered 0.22 deg$^2$ and the cycle 9 observations covered 0.23 deg$^2$. This represents $\sim$75\% and $\sim$65\% of the \nus cycles 8 and 9 field, respectively.

\subsection{Data Reduction}
The \xmm data were reduced following previous surveys' prescriptions (\citealt{Brunner2008}, \citealt{Cappelluti2009}, \citealt{Lamassa2016}, and \citetalias{Zhao2024}). Using the \xmm Science Analysis System (SAS)\footnote{\url{https://www.cosmos.esa.int/web/xmm-newton/sas-thread-src-find-stepbystep}}, we generated the observational data files (ODF) for the three \xmm instruments (MOS1, MOS2, and PN) using the \texttt{emproc} and \texttt{epproc} tasks in SAS version 21.0.0. Time intervals for the MOS and PN files were removed when the background count rates exceeded 0.2 and 0.3 cts/s, respectively. As stated above, $\sim$40\% of the total exposure time in cycles 8 and 9 was removed due to high background levels caused by flares. The clean event files were then used to create images from the MOS1, MOS2, and PN data in the 0.5$-$2\,keV and 2$-10$\,keV bands. \\
\indent The exposure maps in both energy bands were created using the SAS task \texttt{eexpmap}. In order to sum the exposure map of the three instruments, we used energy conversion factors (ECF) for each instrument. We calculated this in WebPIMMs using an absorbed power-law model with $\Gamma$=1.80 and the galactic N$_{\rm H}$ = 3.4 $\times$ 10$^{20}$ cm$^{-2}$. For cycle 8, the MOS and PN ECFs were 1.86 and 6.73 in the \zetw band and 0.45 and 1.18 in the \twte band. For cycle 9, the MOS and PN ECFs were 1.95 and 7.13 in the \zetw band and 0.45 and 1.26 in the \twte band. All ECFs are in units of $\times$ 10$^{-11}$ erg cm$^{-2}$ counts$^{-1}$. \\
\indent In order to produce the background maps for all three instruments, the potential sources in the FoV must be masked. The SAS task \texttt{eboxdetect} was used to perform preliminary source detection using a sliding box detection with the detection likelihood LIKE$>$4 to prevent any possible sources from contaminating the background. Here, the detection likelihood is defined as LIKE = $-$ ln $p$, where $p$ is the probability of a Poissonian random fluctuation producing the counts detected in the detection box. The final step uses \texttt{esplinemap} to create the background map assuming a model containing two components: background from the detector (particle) and unresolved sources.

\subsection{Source Detection}

\begin{figure}
    %\centering
    \hspace{-1.5cm}
    \includegraphics[scale=0.47, clip=true,trim=75 0 0 0]{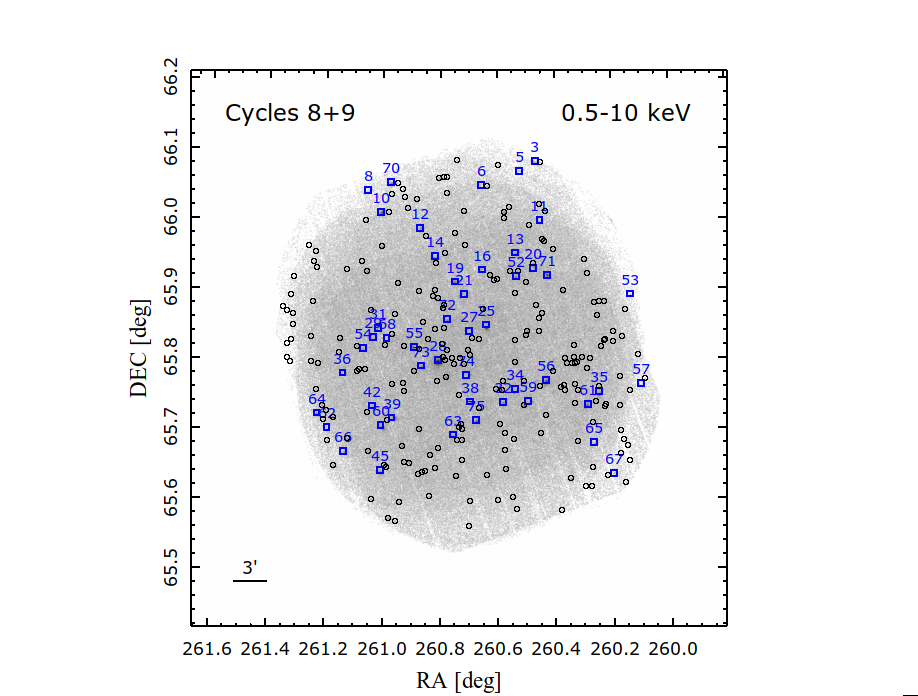}
    \caption{The combined mosaic of the eight \xmm observations from Cycles 8 and 9 in the 0.5$-$10\,keV band, including the 274 sources detected. The blue squares represent the 48 sources with \nus counterparts in cycles 8+9. The associated \nus source IDs are also shown. Figure \ref{fig:xmm_05-10_no_regs} shows this image without regions to make the sources more visible.}
    \label{fig:xmm_05-10}
\end{figure}

Following \citetalias{Zhao2024}, in order to maximize the sensitivity of this survey, we co-added the cleaned images, exposure maps, and simulated background maps of all three instruments into mosaic images using the SAS task \texttt{emosaic}. Figure \ref{fig:xmm_05-10} displays the 0.5$-$10\,keV merged image mosaic of all eight epochs from cycles 8 and 9. The SAS tasks \texttt{eboxdetect} and \texttt{emldetect} were used to carry out the source detection. \texttt{elmdetect} was used to locate the most likely center of the detected source. We performed source detection in both the \zetw and \twte energy bands. Only sources with maximum likelihood \texttt{mlmin}$>$6 were considered real detections. This threshold corresponds to a 97.3\% detection reliability in the \zetw band and a 99.5\% reliability in the \twte band. These numbers were found in the XMM COSMOS survey \cite{Cappelluti2007} which had a similar exposure time to each cycle discussed here ($\sim$60\,ks). Sources detected along the edges of the FoV with a total exposure $<$1\,ks were excluded.

\subsection{Astrometric Correction} \label{sec:astr_cor}

In order to maximize the positional accuracy, we performed astrometric corrections before merging all observations into one mosaic. The average astrometric offset of an \xmm observation is around 1.0$-$1.5$\arcsec$ and rarely exceeds 3$\arcsec$ \citep{Cappelluti2007, Ni2021}. To determine the necessary correction for each observation, we matched our \xmm sources detected at $>$6$\sigma$ (\texttt{mlmin} $>$ 20) with sources from the Sloan Digital Sky Survey (SDSS) DR16\footnote{\url{https://www.sdss4.org/dr16/}}. We only included sources labeled as stars in the DR16 and we used a matching radius of 4$\arcsec$. Each observation in cycles 8 and 9 had $\sim$13 matches and the typical offset in RA and Dec were ($\Delta\alpha, \Delta\delta$) = (1.68$\arcsec$, 0.71$\arcsec$). These corrections were applied to all the event and attitude files and then we remade the images, background maps, and exposure maps using these new files. \\
\indent Once the source detection was completed with the corrected files, we performed the same crossmatch with the SDSS catalog to confirm the offsets had decreased. The corrected average offsets from the DR16 sources decreased by 80\% and 50\%, for RA and DEC, respectively. These images, background maps, and exposure maps were then combined into mosaics. The final average offset between X-ray and SDSS is 0.51$\arcsec$, which we designate as the systematic positional uncertainty in the \xmm NEP TDF survey for cycles 8 and 9. We note that the value found in \citetalias{Zhao2024} for the cycle 6 survey was 1.22$\arcsec$.

\subsection{Sensitivity}
\begin{figure}
    \centering
    \includegraphics[width=1\linewidth]{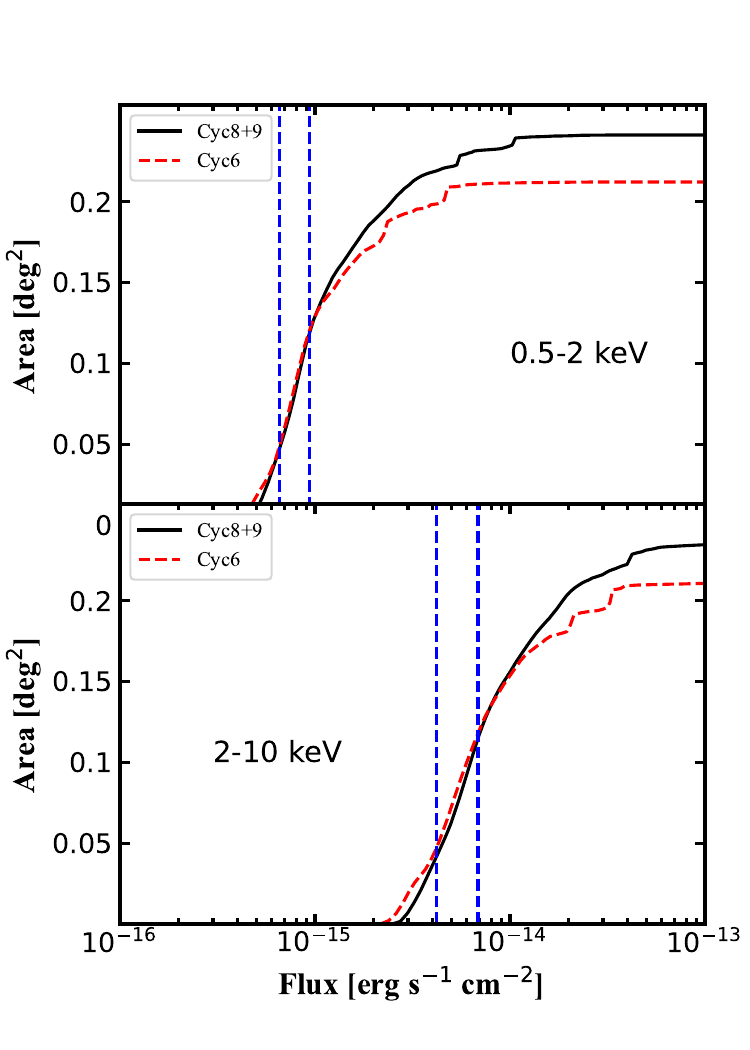}
    \caption{The \zetw (top) and \twte (bottom) sensitivity curves for the \xmm cycles 6 and 8+9 surveys. The blue vertical lines represent the half-area and 20\% area sensitivities for the combined cycles 8+9 survey.}
    \label{fig:xmm_sens}
\end{figure}

Figure \ref{fig:xmm_sens} shows the sensitivity curves for the \xmm cycle 6 and 8+9 surveys. These curves were calculated using the SAS tool \texttt{esensmap} while assuming a maximum likelihood value of \texttt{mlmin} $>$ 6. The values for the half-area and 20\%-area sensitivities for \xmm cycles 8+9 are listed in Table \ref{tab:half_area_flux}.

\subsection{XMM-Newton Source Catalog}

\begin{figure}
    \centering
    \includegraphics[width=1\linewidth]{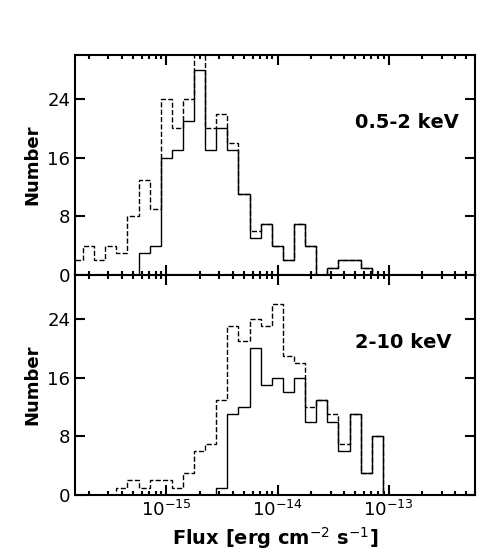}
    \caption{The flux distributions of the sources detected in the cycles 8+9 \xmm survey. The top shows the \zetw band and the bottom shows the sources detected in the \twte band. The solid black line represents the sources detected with \texttt{mlmin} $>$ 6. The dashed line represents the flux at a 90\% confidence upper limit of all 274 detected sources, including those with \texttt{mlmin}$<$6 in that band.}
    \label{fig:flux_distr_xmm}
\end{figure}

The combined \xmm catalog for cycles 8+9 contains 190 sources in the \zetw band and 179 sources in the \twte band. The two bands share 91 sources, while the \zetw band has 99 distinct sources and the \twte band has 88 distinct sources. Therefore, there are a total of 274 sources detected in at least one band. The properties of each source can be found in the public catalog, while the description of each column in the catalog is listed in Table \ref{Table:xmm_cat_info}. Figure \ref{fig:flux_distr_xmm} shows the source flux distributions in both the \zetw and \twte bands.

\subsection{Crossmatch with \nus} \label{sec:xmm_nus_crossmatch}
Following the procedure in \citetalias{Zhao2024}, we crossmatched the 274 \xmm sources with the 75 \nus sources detected in cycles 8+9. This was done using a crossmatch radius of 20" (the \nus positional uncertainty) combined in quadrature with the \xmm source positional uncertainty ($\sigma_{XMM}$ found with \texttt{elmdetect}) and the \xmm systematic uncertainty found above in Section \ref{sec:astr_cor} (0.51$\arcsec$). We found 48 \nus sources with one \xmm counterpart. Unlike \citetalias{Zhao2024}, there were no \nus sources with multiple \xmm counterparts found within the search radius. This left 27 \nus sources without an XMM counterpart from the cycles 8+9 survey.\\

\subsubsection{\xmm Cycle 6 Crossmatch with NuSTAR}
\indent We crossmatched these 27 \nus sources with the \xmm cycle 6 catalog using the XMM source positional uncertainty from \citetalias{Zhao2024} (1.22$\arcsec$). Five of the 27 sources have positional matches. Therefore, 53 \nus sources from cycles 8+9 have \xmm counterparts. \\
\indent Nus$_{89}$\_ID = 24 was detected by \nus above the 99\% reliability level in the \thei band during cycles 8+9 but was not detected by XMM in cycles 8 and 9. However, it was detected by XMM during cycle 6. This suggests potential soft X-ray variability and will be studied further in a future work. \\
\indent Two sources (Nus$_{89}$\_ID = 15 and 26) were detected in the \thei band in the cycles 8+9 \nus data, but below the 99\% reliability threshold. They were also detected by XMM in cycle 6, but not in cycles 8+9. This means they were bright in the soft X-rays during cycle 6 but not during cycles 8 and 9. These sources will also be analyzed in our variability paper. \\
\indent The final two sources (Nus$_{89}$\_ID = 47 and 49) were outside the FoV of the cycles 8+9 \xmm observations (see Fig. \ref{fig:cyc89_mosaic}), but were observed and detected during the \xmm cycle 6 observations. 

\subsubsection{Flux Comparisons}
\begin{figure}
    \centering
    \includegraphics[width=1\linewidth]{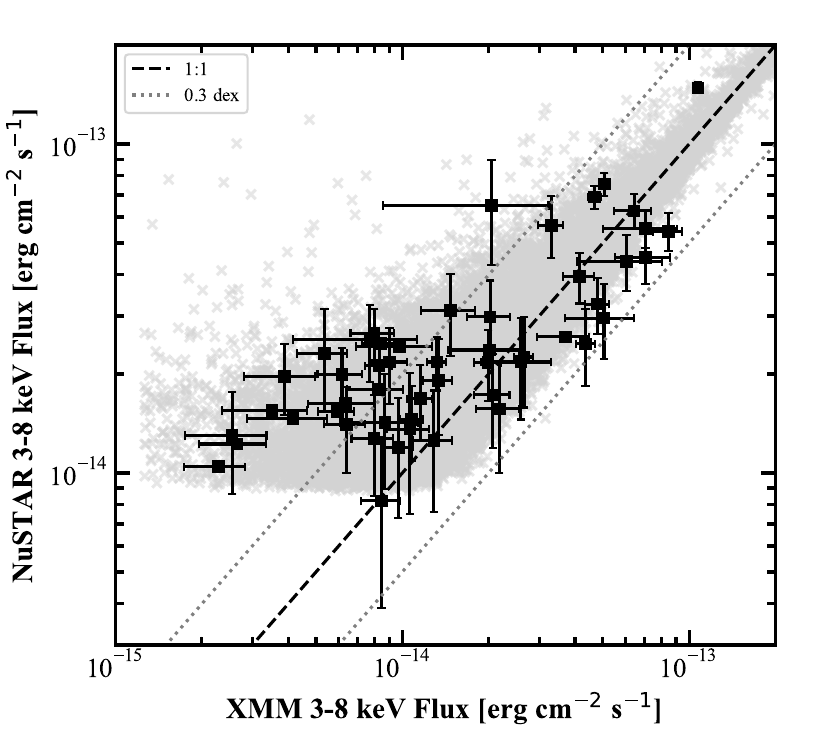}
    \caption{The \nus \thei fluxes compared with the \thei \xmm fluxes of the 53 sources with detections by both instruments in the Cycles 8+9 survey. The \xmm fluxes have been converted from 2$-$10\,keV fluxes using the CF of 0.62 found by \citetalias{Zhao2024}. The black dashed line represents a 1:1 correlation while the grey dotted lines represent a factor of two difference. The grey crosses in the background are taken from the \thei \nus simulations (see Figure \ref{fig:sims_flux} as an example).}
    \label{fig:nus_xmm_38flux}
\end{figure}

\indent Figure \ref{fig:nus_xmm_38flux} shows the comparison of the \nus and \xmm \thei fluxes for the 53 sources detected in both surveys. The \twte \xmm fluxes were converted to \thei fluxes using CF $=$ 0.62, as in \citetalias{Zhao2024}. Most sources have comparable fluxes, particularly when considering uncertainties. We note that \nus fluxes being greater than the \xmm fluxes at the lower end is likely caused by the Eddington bias, as stated in Section \ref{sec:nus_fluxes}. Additionally, these discrepancies could be caused by variability between the \nus and \xmm observations, or differences in spectral shape compared to the simple absorbed power law used to obtain fluxes (see \citetalias{Zhao2024}, Section 5.6). \\

\subsubsection{No XMM Counterparts}
Out of the 75 \nus sources detected in cycles 8 and 9, 22 do not have a soft X-ray counterpart detected by XMM. Three sources, Nus$_{89}$\_ID = 2, 4, and 44, were outside the FoV of any \xmm observation (see Fig. \ref{fig:cyc89_mosaic}). This leaves 19 sources within the footprint of at least one XMM observation but without an XMM detection. \\
\indent Five of these 19 sources were detected above the 99\% reliability level in the \thei band in the cycles 8+9 \nus observations. It is possible they are not detected by XMM due to flux variability. As can be seen in Table \ref{tab:obs_info}, there is only one \xmm observation per \nus epoch. These sources may have been brighter in the \thei band on the days when \nus was observing but fainter when \xmm was observing them. Our future work focused exclusively on variability will analyze these sources in more depth to determine if these sources are variable, obscured, or potentially spurious detections. \\
\indent 11 sources were detected in the \thei band, but below the 99\% reliability level. The final three sources were not detected at all in the \thei band by \nus during cycles 8 and 9, therefore likely do not emit strongly in the soft X-rays.

\section{Multiwavelength Catalogs} \label{sec:mw_sec6}
\begin{figure*}
    \centering
    \includegraphics[scale=1.0, clip=true,trim=50 300 175 0]{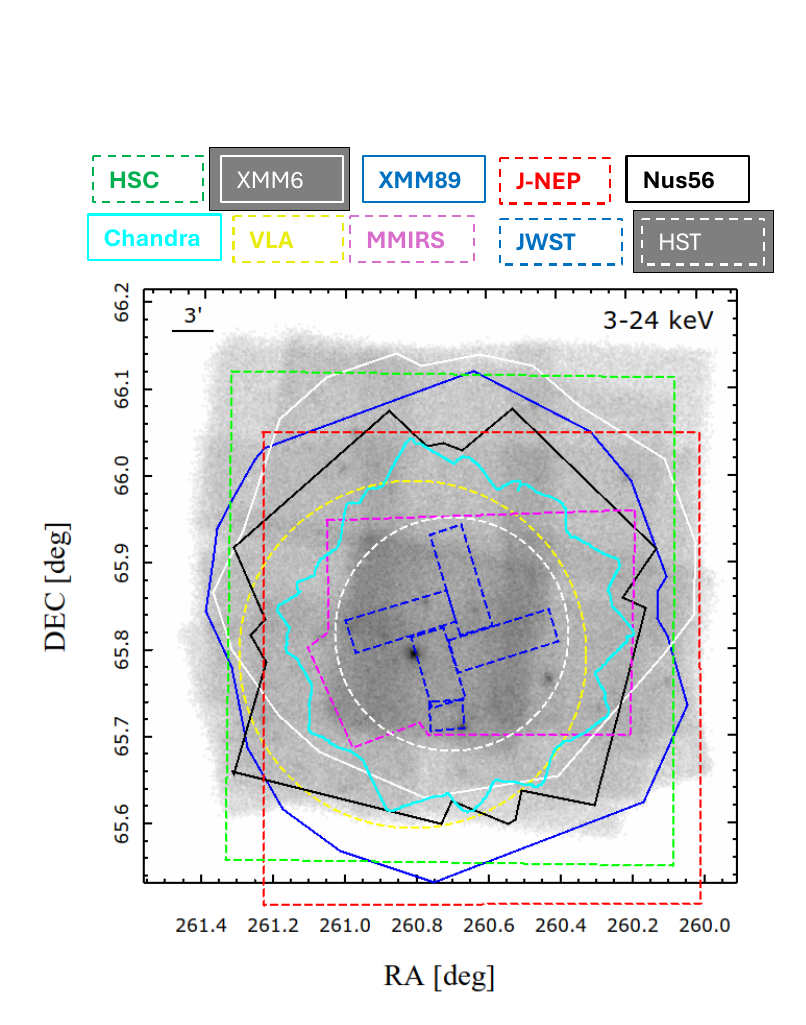}
    \caption{The \nus cycles 8+9 mosaic image in the \thtw band with the areas of all the multiwavelength surveys plotted on top. The \textit{HST} and \textit{JWST} footprints are represented with white and blue dotted outlines, while the XMM cycle 6 and XMM cycles 8+9 are represented with solid white and blue outlines, respectively.}
    \label{fig:MW_regions}
\end{figure*}

The NEP TDF possesses a rich supply of multiwavelength data from all across the electromagnetic spectrum (see Figure \ref{fig:MW_regions}). We crossmatched these multiwavelength catalogs with the sources detected in the cycles 8+9 \xmm catalog.

\subsection{Matching Procedure}
We used a maximum likelihood estimator \citep[MLE;][]{Sutherland1992} that takes into account the physical properties of sources to determine the most likely counterparts, as opposed to solely relying on positional separation. This method was used in previous \xmm and \textit{Chandra} extragalactic surveys and yielded a reliability $>$90\% \citep[see e.g.,][]{Brusa2007, Civano2012, Lamassa2016, Marchesi2016}. First, we crossmatched the XMM sources with the catalog of interest using a 5$\arcsec$ search radius. Using simulations based on the Stripe 82 \xmm survey, \cite{Lamassa2016} demonstrated that greater than 95\% of the XMM sources will be detected within this radius. Next, in order to determine the probability that the matched source is the true counterpart, the MLE method uses 1) the flux and offset of the candidate counterpart, and 2) the positional uncertainties and flux distribution of the survey. The likelihood ratio (LR) can be visualized with the equation:

\begin{equation}
    LR = \frac{q(m)f(r)}{n(m)},
\end{equation}

where \textit{m} is the catalog magnitude of the candidate counterpart, \textit{n(m)} is the local magnitude distribution of background sources near the source in question, \textit{q(m)} is the predicted magnitude distribution of the true multiwavelength counterparts, and finally, \textit{r} is the positional separation between the candidate counterpart and the X-ray source. We measured \textit{n(m)} using an annulus with radii 5$\arcsec$ and 30$\arcsec$ centered on the X-ray source. \textit{q(m)} was set as the normalization of \textit{q\textquotesingle(m)}, i.e., the magnitude distribution of catalog sources within 5$\arcsec$ of the X-ray source, after \textit{n(m)} was subtracted and rescaled to 5$\arcsec$. An example of \textit{q(m)} can be seen in Figure 1 of \cite{Civano2012}. \\
\indent The function \textit{f(r)} is the probability distribution of positional uncertainties, and is assumed to be a 2D Gaussian in the form \textit{f(r)} $=$ 1 / (2$\pi\sigma^2$) $\times$ exp(-$r^2$ / 2$\sigma^2$). Here $\sigma$ represents the positional uncertainty of the \xmm sources (see Section \ref{sec:astr_cor}) added in quadrature with that of the candidate counterpart. For this second value, we used 0.2$\arcsec$ following \citetalias{Zhao2024}. The number of counterparts found in each survey are listed in Table \ref{tab:xmm_counts}. \\
\indent In order to determine whether a candidate was the true counterpart of the X-ray source, or just a spurious background source within the 5$\arcsec$ search radius, we used the LR threshold (LR$_{th}$). The LR$_{th}$ is dependent on the reliability and completeness of the X-ray sample. Both quantities can be approximated from survey statistics \citep{Civano2012}. The formula for the reliability $R_i$ of an individual candidate \textit{j} is as follows:

\begin{equation}
    R_i = \frac{LR_i}{\sum_{i} (LR)_i + (1 - Q)},
\end{equation}

where \textit{Q} is the fraction of \xmm sources that have at least one potential counterpart. This can be thought of as the ratio of column (4) divided by column (3) in Table \ref{tab:xmm_counts}. In the denominator, the LR is summed over all candidate counterparts for the X-ray source within the search radius. To determine the reliability (R$_{tot}$) for the entire sample, divide the summed reliability for every candidate counterpart by the total number of sources with LR $>$ LR$_{th}$. To determine the completeness (\textit{C}) of the sample, divide the summed reliability of the entire sample of candidate counterparts by the number of X-ray sources with potential counterparts within the search radius. \\
\indent As LR$_{th}$ increases, the reliability of the matching also increases, however the completeness decreases. The inverse is true if LR$_{th}$ is decreased. Following \citetalias{Zhao2024} and \cite{Brusa2007}, we selected an LR$_{th}$ by maximizing the quantity (\textit{R} + \textit{C})/2. After incrementally increasing the LR$_{th}$ by 0.1 from 0.1 to 1.0, we found the optimal value for each of the four catalogs (HSC, SDSS, MMIRS, and WISE) to be: 0.4, 0.1, 0.3, and 0.1, respectively. Table \ref{tab:xmm_counts} lists the number of \xmm sources with at least one counterpart from each of the catalogs (column 4). The HSC and MMIRS catalogs were the primary optical and IR catalogs due to their sensitivity and total coverage of the NEP-TDF. If no counterpart was found from these two, the other catalogs were then searched.  \\
\indent For the final step in this process, we performed a visual inspection of all the potential multiwavelength counterparts for each X-ray source. This allowed us to confirm the validity of our results. After this inspection was completed, the candidate counterparts were sorted into three different categories: \\ \\
1. Secure: there is only one counterpart possessing LR $>$ LR$_{th}$. However, an X-ray source with multiple counterparts with LR $>$ LR$_{th}$ can still be labeled as ``secure'' as long as the LR of the primary source is 4x greater than that of the secondary source. \\
2. Ambiguous: the X-ray source has multiple candidate counterparts with LR $>$ LR$_{th}$ and the primary candidate LR is less than 4x greater than that of the secondary source. The counterparts can also be listed as ``ambiguous'' if the secure optical counterpart is different than the secure infrared counterpart. \\
3. Unidentified: the X-ray source does not have any optical or infrared counterparts within the search radius with LR $>$ LR$_{th}$. 
\\ \\
\indent Of the 274 sources in the cycles 8+9 \xmm catalog, 221 were found to have at least one secure counterpart.

\begin{table}[]
    \centering
    \begin{tabular}{cccccc}
        \hline \hline
         Survey & Band & XMM & Matched & Candidates & CP \\
         (1) & (2) & (3) & (4) & (5) & (6) \\
         \hline 
         HSC & \textit{i} & 273 & 245 & 504 & 182 \\
         SDSS & \textit{i} & 274 & 103 & 103 & 101 \\
         MMIRS & \textit{J} & 135 & 88 & 165 & 103 \\
         WISE & W1 & 274 & 207 & 318 & 195 \\
         HST & F606W & 92 & 89 & 1630 & 50 \\
         JWST & F444W & 36 & 35 & 389 & 24 \\
         \hline
    \end{tabular}
    \caption{The \xmm counterparts found in each catalog and which band was used. Column 3) displays the total number of cycles 8+9 \xmm sources covered by the listed catalog. Column 4) lists the amount of XMM sources with at least one potential counterpart within a 5$\arcsec$ radius. Column 5) lists the total amount of possible counterparts within 5$\arcsec$ of an XMM source. Column 6) displays the number of XMM sources with at least one counterpart within 5$\arcsec$ and has an LR$>$LR$_{th}$.}
    \label{tab:xmm_counts}
\end{table}

\iffalse
\begin{minipage}{\linewidth}
\centering
\captionof{table}{Table Title} \label{tab:title} 
\begin{tabular}{ C{1.25in} C{.85in} *4{C{.75in}}}\toprule[1.5pt]
\bf Variable Name & \bf Regression 1 & \bf Mean & \bf Std. Dev & \bf Min & \bf Max\\\midrule
text        &  text     & text      &  text     &  text     &text\\
\bottomrule[1.25pt]
\end {tabular}\par
\bigskip
Should be a caption
\end{minipage}
\fi

\subsection{Optical Catalogs}
Following the procedure in \citetalias{Zhao2024}, we crossmatched our XMM sources with two different optical catalogs that surveyed the NEP TDF: the SDSS DR17 \citep{Abdurro2022} and the HEROES catalog \citep{Taylor2023} comprised of data taken by Hyper Suprime-Cam attached to the Subaru telescope \citep[HSC;][]{Aihara2018}. In both cases, the \textit{i}-band catalog was used as it contains the most detected sources. \\
\indent The SDSS survey covers the entire NEP TDF and the data can be obtained from their public database\footnote{\url{https://www.sdss4.org/dr17/}}. The HSC NEP-TDF catalog \citep{Willmer2023} was derived from the reduction by S. Kikuta of HSC data that were publicly accessible in 2020. We note that sources in this catalog with $m_i <$ 17.5 are saturated and therefore unusable. Thus, they were replaced with the \textit{i}-mags listed in the SDSS catalog. The J-NEP survey \citep{Heran-Caballero2023} was carried out with the 2.55\,m Javalambre Survey Telescope using the SDSS \textit{u, g, r,} and \textit{i} filters. This survey observed approximately $\sim$90\% of the cycles 8+9 \xmm survey. We restricted the catalogs to only sources with S/N $>$ 3 in order to ensure reliable detections and flux measurements. To achieve this, we implemented magnitude cutoffs of $m_i$ $\leq$ 22.5, 25.8, and 24.5 for the SDSS, HSC, and J-NEP catalogs, respectively. \\
\indent In addition to these two catalogs, we also crossmatched with the optical catalog from HST \citep{Obrien2024}. For more details, see Section \ref{sec:hst_jwst}.

\subsection{Infrared Catalogs}
The two near-IR catalogs used for counterpart identification (in addition to JWST; see Section \ref{sec:hst_jwst}) are the 1) YJHK catalog \citep{Willmer2023} comprised of data taken by the MMT-Magellan Infrared Imager and Spectrometer \citep[MMIRS;][]{Mcleod2012}, and the 2) unWISE catalog \citep{Schlafly2019} which utilizes 5 years of data taken by the Wide-field Infrared Survey Explorer \citep[WISE;][]{Wright2010}. This catalog covers two wavebands: 3.4\,$\mu$m (W1) and 4.6\,$\mu$m (W2). \\
\indent The MMIRS catalog observed roughly 30\% of the cycles 8+9 \xmm catalog, while the unWISE catalog covered the entire survey. To maintain the limit of S/N $>$ 3 mentioned above, we imposed AB magnitude cuts for the MMIRS catalog at $mag \leq$ 24.6, 24.5, 24.1, and 23.5 (corresponding to the \textit{Y, J, H,} and \textit{K} bands, respectively) and cuts of $mag \leq$ 21.5 and 20.5 for the W1 and W2 filters from the unWISE catalog.

\subsection{Radio Counterparts}
In addition to optical and infrared surveys, the NEP-TDF was observed in the radio by the Karl G. Janksy VLA (PIs: R.A. Windhorst \& W. Cotton). This VLA survey utilized the ``S band$\arcsec$ ($\nu$ = 3\,GHz) and was granted 48 hours of exposure time \citep{Hyun2023}. It covered $\sim$0.13\,deg$^2$, approximately half of the \xmm survey, and was centered on the blazar ($z$ = 1.441; Nus$_{56}$\_ID = 29, Nus$_{89}$\_ID = 28). The NEP-TDF VLA survey contains 756 sources with S/N $>$ 5 and an angular resolution of FWHM = 0.7$\arcsec$. Using this as the search radius in conjunction with the XMM positional uncertainty, we discovered 65 matches out of the 274 sources in the cycles 8+9 \xmm survey. This is a similar match percentage to what was found in \citetalias{Zhao2024} and COSMOS, where both found $\sim$30-40\% of their X-ray sources had VLA counterparts \citep{Marchesi2016, Smolvcic2017}. Of the 65 VLA counterparts found, the brightest source is the blazar with a 3\,GHz flux of 0.2\,Jy, and the median flux of all the counterparts is 30 $\mu$Jy. \\
\indent The \cite{Willner2023} IR survey found 62 VLA sources had JWST counterparts. Of those, 6 also had counterparts in \xmm cycles 8+9 (XMM$_{8+9}$\_ID = 3, 13, 15, 44, 80, 128). Four of these sources were also detected by XMM in cycle 6 and found the same VLA counterpart (XMM$_{8+9}$\_ID = 3, 13, 15, 44). \\ 
\indent \cite{Hyun2023} reported 114 sources with S/N $>$ 3.5 found in the NEP-TDF by the James Clerk Maxwell Telescope (JCMT) SCUBA-2 850\,$\mu$m survey. Of these 114, nine were found to have counterparts in the \xmm cycles 8+9 survey, compared to four XMM sources in the cycles 5+6 survey.

\subsection{HST and JWST Counterparts} \label{sec:hst_jwst}
In addition to the optical surveys mentioned above, HST also observed the NEP-TDF (GO15278, PI: R. Jansen; GO16252/16793, PIs: R. Jansen \& N. Grogin). This program included imaging with the F275W filter on the WFC3/UVIS, and the F435W and F606W filters on the Advanced Camera for Surveys (ACS)/WFC \citep[][Jansen et al. in preparation]{Obrien2024}. Covering an area of $\sim$194\,arcmin$^2$, these observations reached a 2$\sigma$ limiting depth of mag$_{AB} \simeq$ 28.0, 28.6, and 29.5 in the F275W, F435W, and F606W filters, respectively. \\ 
\indent In addition to the IR data listed above, JWST observed the NEP-TDF four times between August 2022 and May 2023 (PI:
R. A. Windhorst \& H. B. Hammel; PID 2738)\footnote{These JWST data were obtained from the Mikulski Archive for Space Telescopes (MAST) at the Space Telescope Science Institute. The specific observations analyzed can be accessed via \dataset[DOI: 10.17909/b36q-ct37]{https://doi.org/10.17909/b36q-ct37}.}. The survey utilizes eight NIRCam filters (F090W, F115W, F150W, F200W, F277W, F356W,
F410W, and F444W) with 5$\sigma$ point-source limits of 28.6, 28.8, 28.9, 29.1, 28.8, 28.8, 28.1, and 28.3 AB mag, respectively. Each of the four NIRCam images observed an area of 2.15$\arcmin$ $\times$ 6.36$\arcmin$, adding up to a total coverage of $\sim$55\,arcmin$^2$. To complement the imaging, the survey includes NIRISS grism data with a 1$\sigma$ continuum sensitivity equal to 25.9. The NIRISS epochs covered 2.22$\arcmin$ $\times$ 4.90$\arcmin$. \\
\indent Figure \ref{fig:MW_regions} shows the area of the NEP-TDF covered by each survey. The HST footprint covers 25\% of the \xmm survey, while the JWST footprint only covers $\sim$10\%. Due to the limited coverage of this field, we did not use HST and JWST in the initial searches for XMM counterparts. Instead, we crossmatched the HST and JWST catalogs with the multiwavelength counterparts already found from the catalogs mentioned above. When performing the crossmatching, we used the F606W HST and F444W JWST catalogs, as they were the filters with the deepest sensitivities. We found the optimal LR$_{th}$ to be 1.0 and 0.8 for HST and JWST, respectively. These values yielded 50 XMM sources with secure HST counterparts and 24 XMM sources with secure JWST counterparts. We note that when using catalogs based on other filters, we found the same HST and JWST counterparts. This confirms our primary method was unbiased. \\

\subsection{\textit{Chandra} Counterparts}
In addition to the \xmm data, the NEP TDF also has extensive (1.8\,Ms) soft X-ray coverage from \textit{Chandra}\footnote{This paper employs a list of Chandra datasets, obtained by the \textit{Chandra} X-ray Observatory, contained in the \textit{Chandra} Data Collection ~\dataset[DOI: 10.25574/cdc.475]{https://doi.org/10.25574/cdc.475}.} (Maksym et al. private communication). We performed a crossmatch with radius determined by adding the \nus (20$\arcsec$) and \cha (0.5$\arcsec$) positional uncertainties in quadrature. We found 49 \nus sources from cycles 8+9 with \cha counterparts. %One of these sources, (Cha\_ID = 394, Nus$_{89}$\_ID = 29, XMM$_{89}$\_ID = 225) was able to resolve ambiguous multiwavelength counterparts from WISE and MMIRS. \\
Of these 49, we found 3 \cha sources (Nus$_{89}$\_ID = 18, 23, and 41) that did not have a counterpart from either \textit{XMM} catalog. We followed the same procedure laid out above to find multiwavelength counterparts for these three sources. As the sample was limited, we found the optimal value of LR for every catalog to be 0.1. The results from this search are as follows: \\ \\
- Nus$_{89}$\_ID = 41 was the only source to have an optical counterpart. It was found in both the HSC and SDSS catalogs. \\
- Nus$_{89}$\_ID = 18 was the only source to have an MMIRS counterpart. \\
- Nus$_{89}$\_ID = 23 and 41 were both found in the unWISE catalog. \\
- Nus$_{89}$\_ID = 41 was detected in the 3\,GHz VLA catalog, while Nus$_{89}$\_ID = 18 was detected in the 850\,$\mu$m VLA catalog. \\

%Thus, all three \nus sources were found in at least one lower-energy wavelength. 

\subsection{Comparison with Cycles 5+6 Counterparts}

\subsubsection{XMM Sources}
We performed a crossmatch between the \xmm cycles 6 and 8+9 surveys and determined 107 sources were detected in both surveys. We then proceeded to compare the counterparts found in each survey to verify the validity of our methods. \\
\indent We started with the optical counterparts. Of the 107 sources, 83 had the identical single counterpart from the HSC catalog. One source, the blazar (XMM$_{89}$\_ID = 2), was saturated in the HSC observations and thus we used the SDSS counterpart. 16 XMM sources had no HSC or SDSS counterparts. This leaves seven sources. \\
\indent For two sources (XMM$_{89}$\_ID = 118, 182), the 5+6 counterparts passed some threshold levels, but not the optimal LR$_{th}$=0.4 found in this work for the HSC catalog. For XMM$_{89}$\_ID = 20, both the cycle 6 and 8+9 catalogs found it to have two counterparts above the threshold level. For three sources (XMM$_{89}$\_ID = 53, 104, 115), this work found multiple counterparts above the threshold while \citetalias{Zhao2024} only found one. For the final source, XMM$_{89}$\_ID = 275, the counterpart found in \citetalias{Zhao2024} was not above the threshold for any value of LR$_{th}$. This is the only source for which this is the case. This work, instead, found four HSC sources above the LR$_{th}$ value. \\
\indent We performed these checks on all the other wavebands discussed in this section and achieved a similar level of success matching counterparts with the \citetalias{Zhao2024} catalog. \\

\subsubsection{NuSTAR sources}
We performed the same comparisons with the 23 \nus sources detected in both the cycles 5+6 and 8+9 surveys. 20 sources were found to have the same multiwavelength counterparts. For Nus$_{89}$\_ID = 68, this work found two ambiguous optical counterparts, one of which, was the only secure optical counterpart found in the cycles 5+6 survey (Nus$_{56}$\_ID = 45). The source Nus$_{89}$\_ID = 34 (Nus$_{56}$\_ID = 35) did not have any multiwavelength counterparts in any band for either survey. Finally, the counterparts for Nus$_{89}$\_ID = 59 did not match the counterparts found for its cycles 5+6 counterpart (Nus$_{56}$\_ID = 57). Overall, the multiwavelength counterparts for the \nus sources agree considerably well.

\subsection{Redshifts} \label{sec:redshifts}
\citetalias{Zhao2024} laid out a program that used the Hectospec \citep{Fabricant2005} instrument on the 6.5\,m MMT to obtain optical spectra of the NEP TDF sources (PI: Zhao). In two different runs, a total of 78 sources had spectra obtained with a S/N sufficient to measure the redshift. Of these sources, 47 were also detected in the \xmm cycles 8+9 survey. We found two additional sources in the 8+9 \xmm survey with spectroscopic redshifts from the SDSS DR17. We plotted these sources based on their source classifications in the right panels of Figure \ref{fig:opt_vs_xr}.\\
\indent We also searched the SDSS DR17 catalog for photometric redshifts and found values for an additional 26 sources. Moreover, the crossmatches with JWST contributed another 17 photometric redshifts. In total, out of the 274 \xmm sources from cycles 8+9, 92 have a redshift value. \\
\indent We note that we performed a crossmatch with the HST NEP TDF sources as well, but this did not yield any new sources with redshift values not in hand.

\subsection{X-Ray vs Optical Properties}

\begin{figure*}
    \centering
    \includegraphics[width=1.0\linewidth]{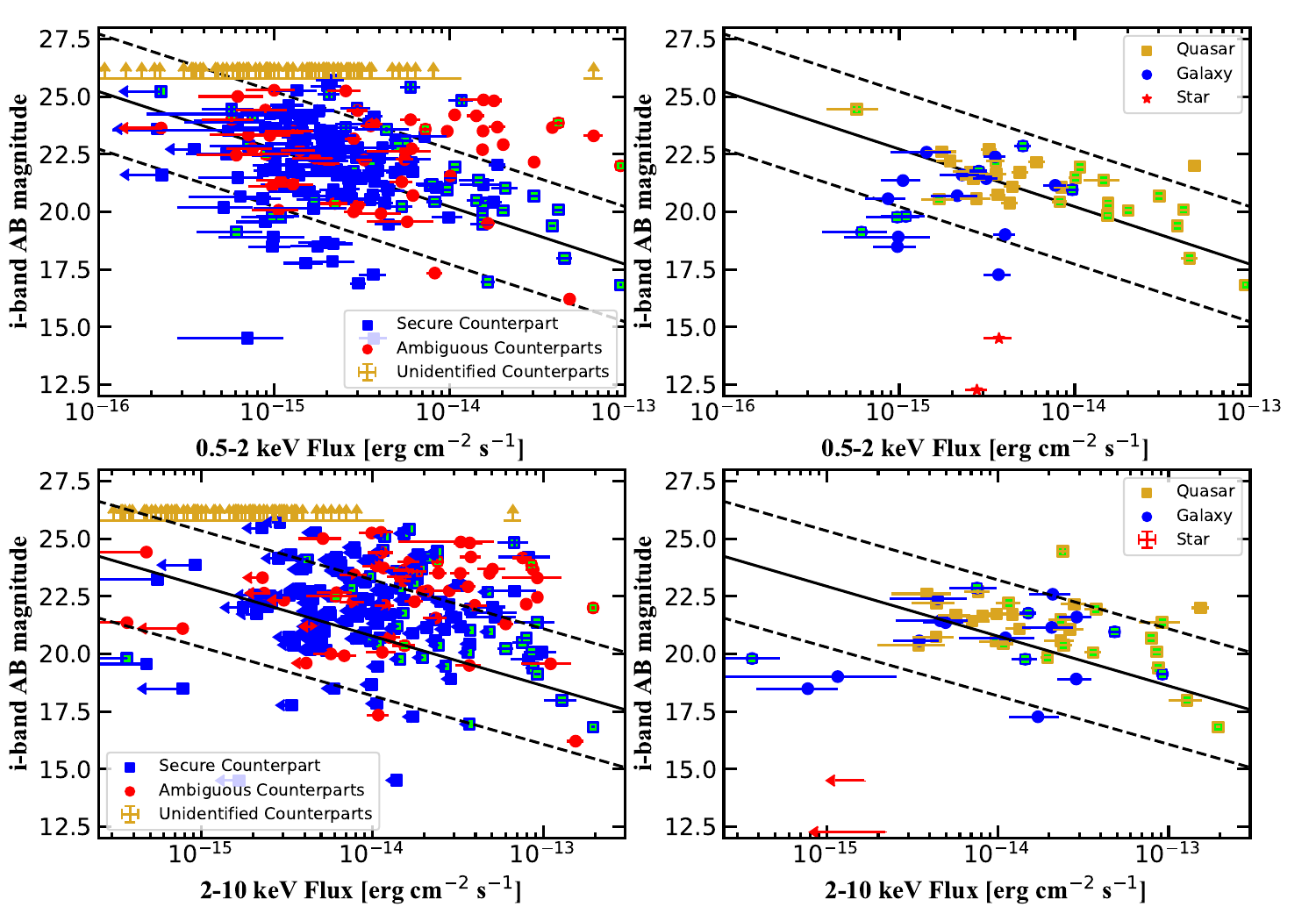}
    \caption{The four plots compare the optical and X-ray properties of the \xmm cycles 8+9 sources with confirmed counterparts. The top row shows the \textit{i}-band magnitudes versus the 0.5$-$2\,keV fluxes while the bottom rows shows the \textit{i}-band magnitudes versus the 2$-$10\,keV fluxes. In the left column, the blue squares represent sources with only one secure optical counterpart above the LR threshold, the red circles represent XMM sources with ambiguous counterparts, and the gold arrows are X-rays sources without an optical counterpart (lower limit magnitude of \textit{i}=25.8). In the right column, the gold squares are sources optically classified as quasars, the blue circles are classified as galaxies, and the red stars as stars. In all four plots, the green points are XMM sources with \nus counterparts from the cycles 8+9 catalog. The solid and dashed lines represent the typical AGN parameter space, $X/O = 0 \pm 1$ \citep{Maccacaro1988}.}
    \label{fig:opt_vs_xr}
\end{figure*}

\begin{figure}
    \centering
    \includegraphics[width=1.0\linewidth]{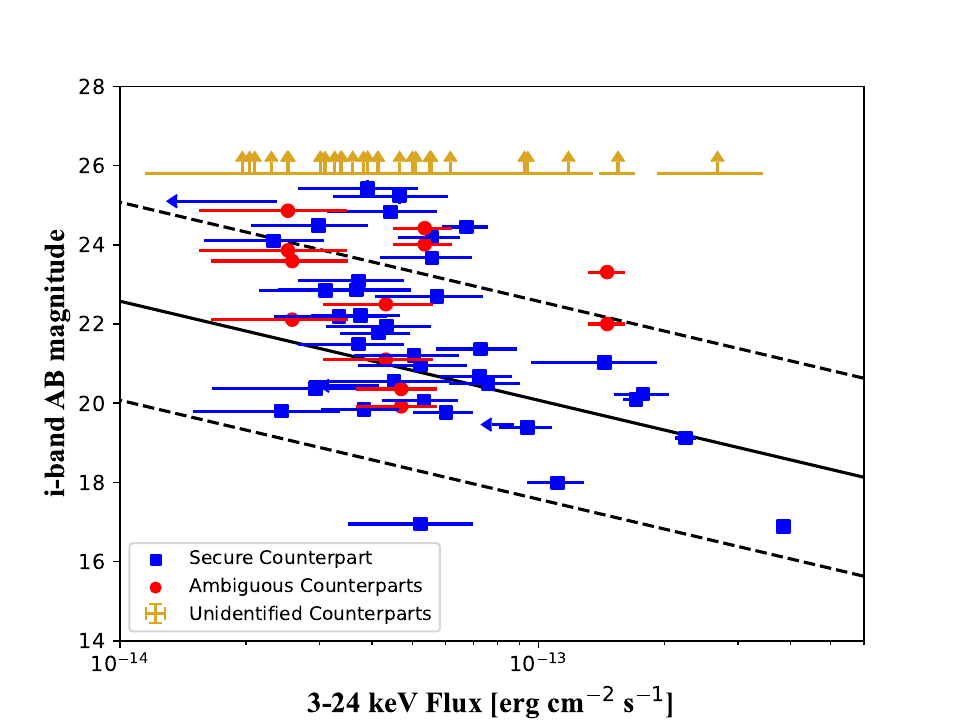}
    \caption{The optical i-band magnitudes versus the \thtw \nus fluxes for the 45 sources from cycles 8+9 that have optical counterparts. 30 sources did not have an optical counterpart and are shown as gold lower limits set to $i$ = 25.8. As in the above plot, the solid and dashed lines represent $X/O = 0 \pm 1$.}
    \label{fig:nus_vs_opt}
\end{figure}

Since the work of \cite{Maccacaro1988}, the X-ray-to-optical flux ratio ($X/O$) has been used to identify the true nature of X-ray sources. This ratio was defined as:

\begin{equation} \label{eq:xr_vs_opt}
   X/0 \equiv log(f_X / f_{opt}) = log(f_X) + m_{opt}/2.5 + C,
\end{equation}

where $f_X$ is the X-ray flux in units of erg cm$^{-2}$ s$^{-1}$, $m_{opt}$ is the optical magnitude in the AB system, and C is a constant dependent on the selected X-ray and optical bands. Figure \ref{fig:opt_vs_xr} shows the $i$-magnitudes versus the \zetw band (top) and the \twte (bottom) for the HSC and SDSS counterparts found for the cycles 8+9 XMM sources. Following the work from \cite{Marchesi2016}, we used the constants $C_{0.5-2}$ = 5.91 and $C_{2-10}$ = 5.44. The left plots show the XMM sources with a secure optical counterpart (blue squares), ambiguous optical counterparts (red circles), or unidentified counterparts (gold lower limits at $i$-mag = 25.8). The right plots only show the sources optically classified as quasars (gold squares), galaxies (blue circles), or stars (red stars). The sources filled with green possess a \nus counterpart from the cycles 8+9 survey.\\
\indent As can be seen in all four plots, the majority of our sample falls in the range -1 $< X/0 <$ 1. This is consistent with previous X-ray surveys \citep{Stocke1991, Schmidt1998, Akiyama2000, Marchesi2016, Zhao2024}. Such works have shown that sources with $X/O >$1 are likely to be at high redshifts or heavily obscured, while sources with $X/O <$1 are more likely to be stars. This supported by the right plots in Figure \ref{fig:opt_vs_xr}. \\
\indent Figure \ref{fig:nus_vs_opt} shows the i-mag versus \thtw fluxes for the 45 cycles 8+9 \nus sources with an optical counterpart. The constant C used here to satisfy equation \ref{eq:xr_vs_opt} was C$_{3-24}$ = 4.97. This plot follows the trend seen in Figure \ref{fig:opt_vs_xr} where the vast majority of sources fall within the -1 $< X/0 <$ 1 region.

\subsection{Luminosity-Redshift Distribution}
\begin{figure*}
    \centering
    \includegraphics[width=0.5\linewidth]{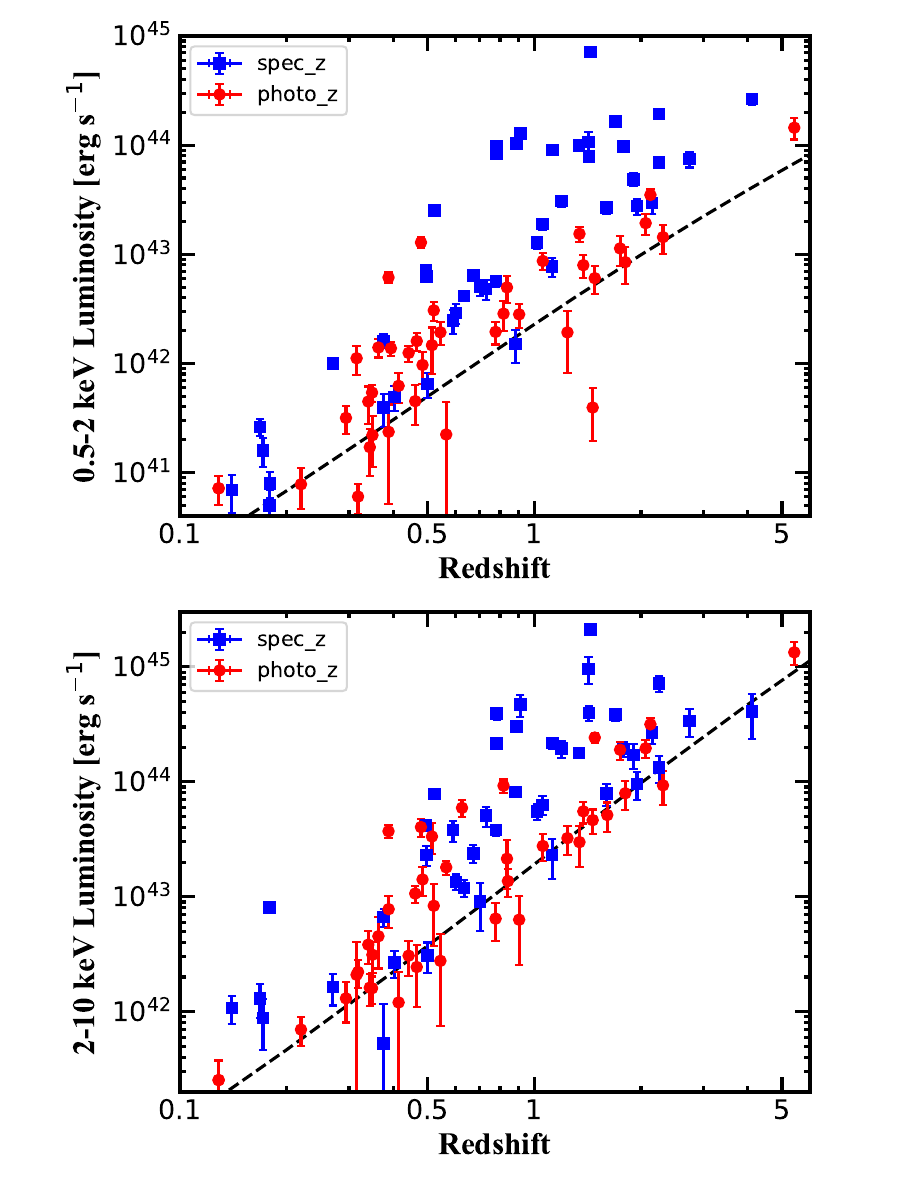}
    \caption{The \zetw (top) and \twte (bottom) rest-frame luminosities versus redshift for the 92 \xmm sources from cycles 8+9 with a redshift value from their secure multiwavelength counterpart. The luminosity values are observed, i.e., they have not been corrected for absorption. The blue squares are sources with a spectroscopic redshift and the red circles are sources with a photometric redshift. The 20\%-area sensitivities (see Table \ref{tab:half_area_flux}) are plotted as dashed black lines.}
    \label{fig:XMM_Lum_total}
\end{figure*}

\begin{figure}
    \centering
    \includegraphics[width=1.0\linewidth]{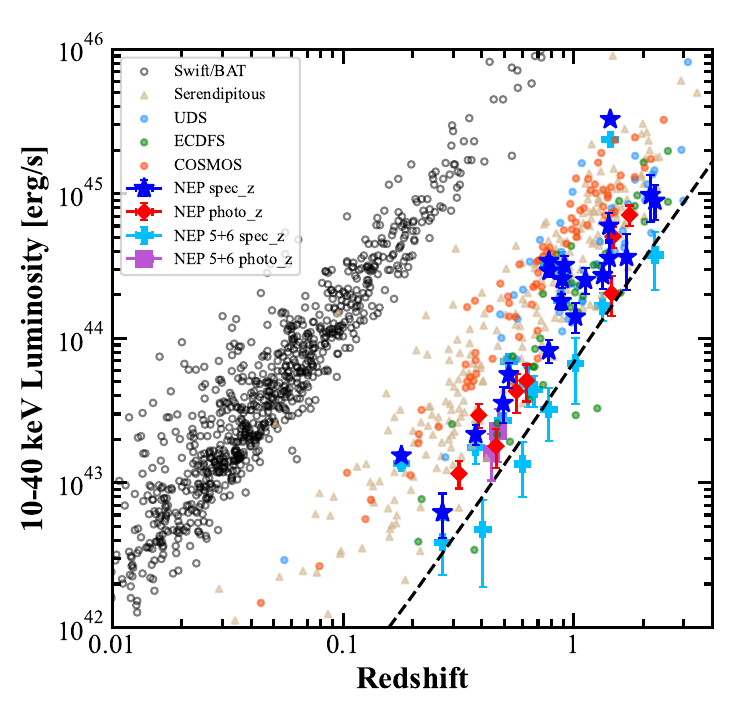}
    \caption{The 10$-$40\,keV rest-frame luminosities of the 29 \nus cycles 8+9 sources with redshift values from their secure multiwavelength counterpart. These luminosity values were extrapolated using the \thtw values. The blue stars are sources with spectroscopic redshifts and the red diamonds are sources with photometric redshifts. The light blue crosses and purple squares represent the sources with spec-z and photo-z, respectively, for the cycles 5+6 \nus sources. The black dotted line represents the 20\%-area sensitivity of the cycles 8+9 \nus survey (see Table \ref{tab:half_area_flux}). The other surveys plotted are as follows: \textit{Swift}-BAT 105 month survey \citep[black open circles;][]{Oh2018}, 40-month Serendipitous survey \citep[gold triangles;][]{Lansbury2017}, UDS \citep[blue circles;][]{Masini2018a}, ECDFS \citep[green circles;][]{Mullaney2015}, and \nus COSMOS \citep[orange circles;][]{Civano2015}. We note that the luminosities in this plot were not corrected for absorption.}
    \label{fig:Lum_10-40}
\end{figure}

Figure \ref{fig:XMM_Lum_total} shows the \zetw (top) and \twte (bottom) rest-frame luminosities plotted against redshift for the 92 cycles 8+9 XMM sources with redshift values. The blue squares represent sources with spectroscopic redshift determinations and the red circles are sources with photometric redshifts (see Section \ref{sec:redshifts}). Following the procedure from \citetalias{Zhao2024}, we calculated the luminosities by converting the observed \zetw and \twte fluxes using a \textit{k}-correction factor. To do this, we used X-ray spectral indices of $\Gamma$ = 1.40 and 1.80 for the \zetw and \twte bands, respectively. We note that the final luminosity values were not absorption-corrected, although using a hard $\Gamma$ of 1.40 in the \zetw band does partially account for it. \\
\indent A similar process was performed for the \nus sources. Figure \ref{fig:Lum_10-40} shows the 10$-$40\,keV rest-frame luminosities of the 29 \nus sources from cycles 8+9 that have redshift values. These were extrapolated from the \thtw fluxes using a \textit{k}-correction with $\Gamma$ = 1.80. The plot also compares the values from this work with that of other high-energy surveys such as: the \textit{Swift}-BAT 105 month survey \citep[black open circles;][]{Oh2018}, 40-month Serendipitous survey \citep[gold triangles;][]{Lansbury2017}, UDS \citep[blue circles;][]{Masini2018a}, ECDFS \citep[green circles;][]{Mullaney2015}, and \nus COSMOS \citep[orange circles;][]{Civano2015}. The largest sample of these comes from the \textit{Swift}-BAT survey which observed the entire sky for 105 months. As this instrument is less sensitive than \textit{NuSTAR}, it is not able to reach as high redshifts. Thus, the median redshift of this sample was $\langle z_{BAT} \rangle$ = 0.044, while the median redshift of the cycles 8+9 \nus survey is $\langle z_{NuS,89} \rangle$ = 0.885. We note that with the addition of the JWST photometric redshifts, this value has increased from the $\langle z_{NuS,56} \rangle$ = 0.734 of the cycles 5+6 \nus survey.

\section{Discussion} \label{sec:disc_sec7}

\subsection{log\textit{N}$-$log\textit{S}} \label{sec:logn-logs}
\citetalias{Zhao2021} first calculated the cumulative number count distribution (log$N-$log$S$) for \nus NEP sources in three energy bands: 3$-$8\,keV, 8$-$16\,keV, and 8$-$24\,keV. Then, \citetalias{Zhao2024} extended this distribution to fainter fluxes due to the increased exposure time. In this work, we compare the \logn calculated from the \nus cycles 5+6 and 8+9 surveys. First, we provide a brief description of how this is calculated. \\
\indent We define the \logn distribution in the same way as \cite{Cappelluti2009}: 

\begin{gather}
     N(>S) \equiv \sum_{i=1}^{N_S} \frac{1}{\Omega_i} deg^{-2},
\end{gather}

where \textit{N($>$S)} is the density of sources that were detected above a 99\% reliability threshold in a given energy band and possess a flux greater than \textit{S}. $\Omega_i$ is the sky coverage associated with the flux of the \textit{i}th source (see Figure \ref{fig:area_coverage} for reference). To calculate the variance of the \textit{N($>$S)}, one must use: 

\begin{gather}
    \sigma^2_S = \sum_{i=1}^{N_S} \left( \frac{1}{\Omega_i} \right)^2.
\end{gather}

\begin{figure}
    \centering
    \includegraphics[width=1\linewidth]{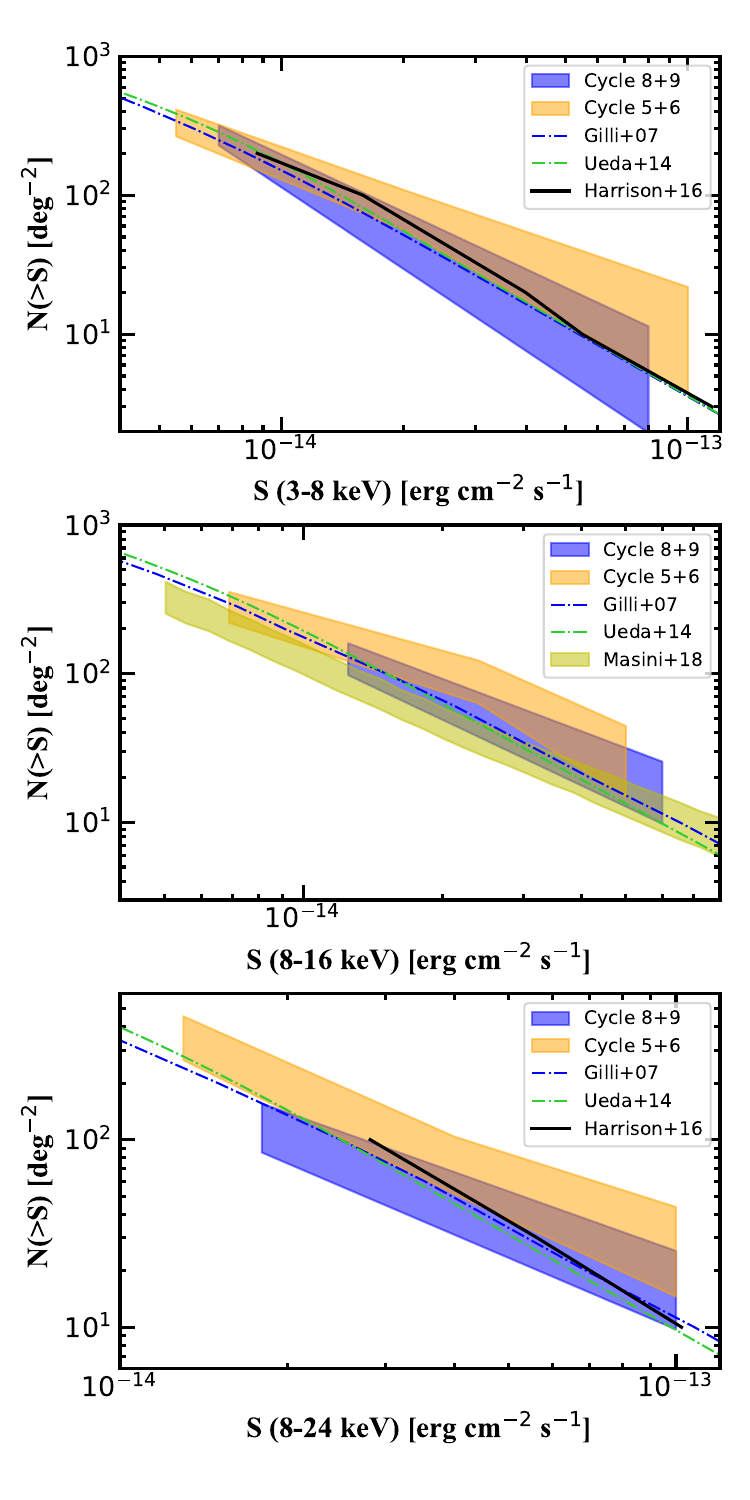}
    \caption{The cumulative source number counts as a function of flux (\logn distributions) in the following energy bands: 3$-$8\,keV, 8$-$16\,keV, and 8$-$24\,keV. The distributions from the cycles 8+9 data are plotted as a blue region while the cycles 5+6 distributions are plotted as an orange region. Both shaded areas represent the 68\% confidence region. In the top panel, the black solid line comes from \cite{Harrison2016}. In the middle panel, the yellow shaded region comes from \cite{Masini2018b}. In all three plots, the dash-dotted lines represent the population synthesis models from \cite{Gilli2007} and \cite{Ueda2014}.}
    \label{fig:lognlogs}
\end{figure}

As shown in \cite{Cappelluti2009} and \cite{Puccetti2009}, the \logn distribution is dependent on the minimum flux limit and the S/N limit of the detected sources. In order to select the proper values for each band, we calculated the \logn of the 1200 simulations (see Section \ref{sec:sims}) with different flux and S/N limits and then compared the results to the input \logn distribution from \cite{Treister2009}. For all three bands, we found the optimal S/N to be 2.5. The optimal flux limit for each band was 7 $\times$ 10$^{-15}$ erg cm$^{-2}$ s$^{-1}$ (3$-$8\,keV), 1.25 $\times$ 10$^{-14}$ erg cm$^{-2}$ s$^{-1}$ (8$-$16\,keV), and 1.8 $\times$ 10$^{-14}$ erg cm$^{-2}$ s$^{-1}$ (8$-$24\,keV). These final values were selected through numerous trials of different configurations. S/N = 2.5 was found to be the best value for all bands because the lowest flux bin in the \thei band was overestimated by $\sim$50\% when using S/N = 2.0. Moreover, using S/N = 4.0 underestimated the lowest flux bin by $\sim$50\%. Similar trends were found with the other two bands. \\
\indent Figure \ref{fig:lognlogs} shows the \logn distributions for cycles 8+9 compared with the cycles 5+6 distributions in all three bands mentioned above. Recent works have compared the slopes of the \logn curves in the \eitw band to the slope expected from a Euclidean geometry ($\alpha=1.50$). \cite{Harrison2016} and \cite{Akylas2019} have found slopes that are steeper than the Euclidean value ($\alpha=1.76 \pm 0.10$ and $\alpha=1.71 \pm 0.20$, respectively), with this difference attributed to the evolution of AGN at higher redshifts. However, \cite{Zappacosta2018} found a flatter slope of  $\alpha=1.36 \pm 0.28$. In this work, we report slopes of $\alpha_{56}=1.24 \pm 0.41$ from the cycles 5+6 curve and  $\alpha_{89}=1.13 \pm 0.46$ from the cycles 8+9 curve. While these slopes are flatter than those from previous works, they are still consistent with the Euclidean value within uncertainties. \\
\indent We believe these curves are flatter than other works due to the presence of the bright blazar (Nus$_{56}$\_ID = 29, Nus$_{89}$\_ID = 28) in this small field of only 0.31\,deg$^2$. This source has increased the N($>$S) value at bright fluxes ($\sim$10$^{-13}$ erg cm$^{-2}$ s$^{-1}$), thus flattening the overall shape of the \logn curve. When the blazar is removed from our sample, the two curves steepen to new slopes of $\alpha_{56}=1.36 \pm 0.46$ and $\alpha_{89}=1.23 \pm 0.50$. Not only are these slopes more consistent with the Euclidean value and \cite{Zappacosta2018}, they are also now consistent with the results from \cite{Harrison2016} and \cite{Akylas2019}.

%While they agree well with each other and the other works plotted overall, there are a few differences present. When comparing the high flux regimes in the \thei and \eitw bands, the cycles 8+9 distributions do not overestimate previous works quite as much as the cycles 5+6 distributions. However, due to the different goal of this survey (area over depth, see Figure \ref{fig:exp_map}), the cycles 8+9 \logn do not reach the same faint flux levels as the 5+6 distributions do. With that said, the 8+9 distribution is still fainter than the one reported in \cite{Harrison2016} in the \eitw band. Interestingly, while the 5+6 data seem to overestimate the number of faint sources in the \eitw band, the 8+9 data appears to underestimate them. Combining these datasets will likely result in a curve that more closely agrees with the population synthesis models of \cite{Gilli2007} and \cite{Ueda2014}.

\subsection{Hardness Ratios} \label{sec:hard_ratios}
\begin{figure*}
    \centering
    \includegraphics[width=0.5\linewidth]{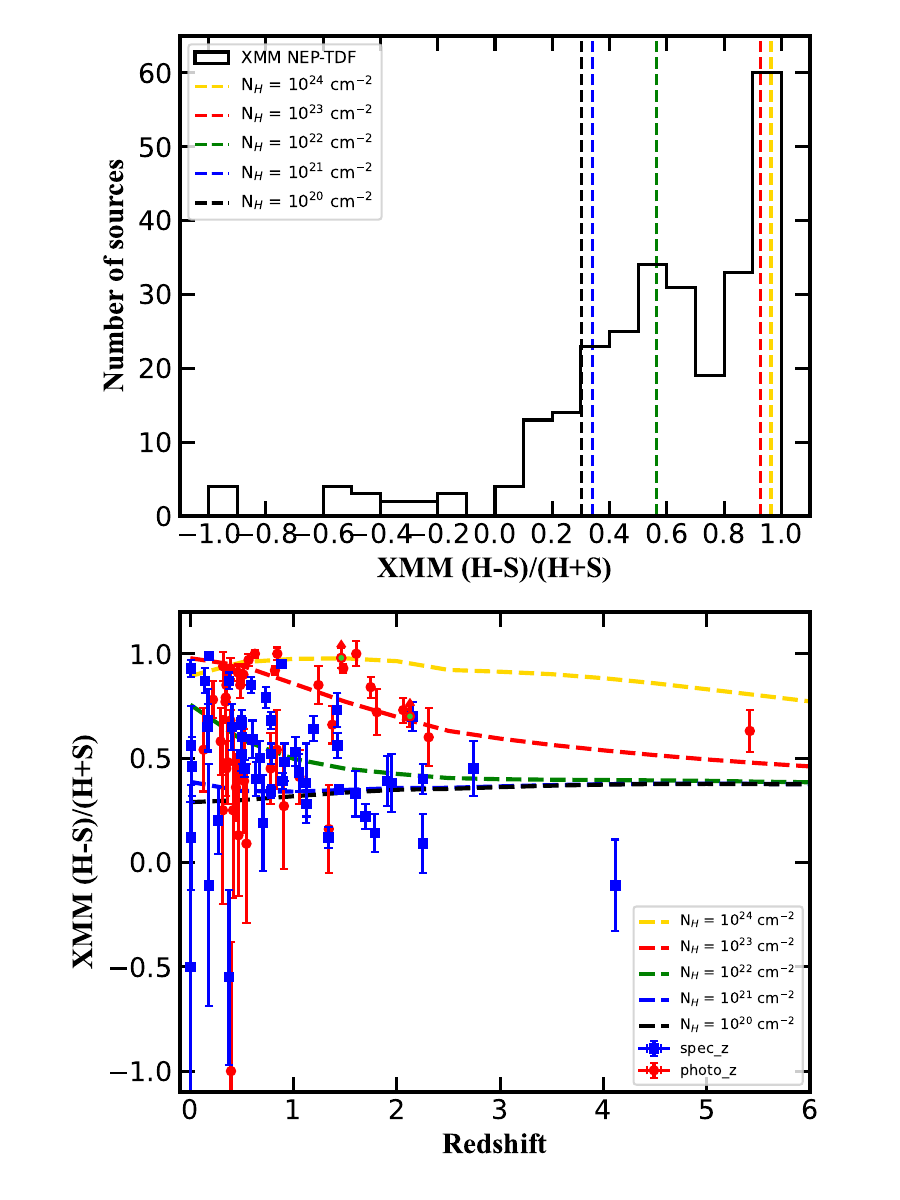}
    \caption{\textbf{Top panel:} The hardness ratios of the 274 \xmm sources detected in cycles 8+9. S and H represent the \zetw and \twte bands, respectively. The vertical dashed lines indicate the likely column density associated with that hardness ratio assuming a redshift of 0. \textbf{Bottom panel:} The HRs vs $z$ for the 92 XMM sources with a known redshift. The two green sources are the \nus sources expected to be CT (see Figure \ref{fig:nus_hr}). }
    \label{fig:xmm_hr}
\end{figure*}

\begin{figure*}
    \centering
    \includegraphics[width=0.5\linewidth]{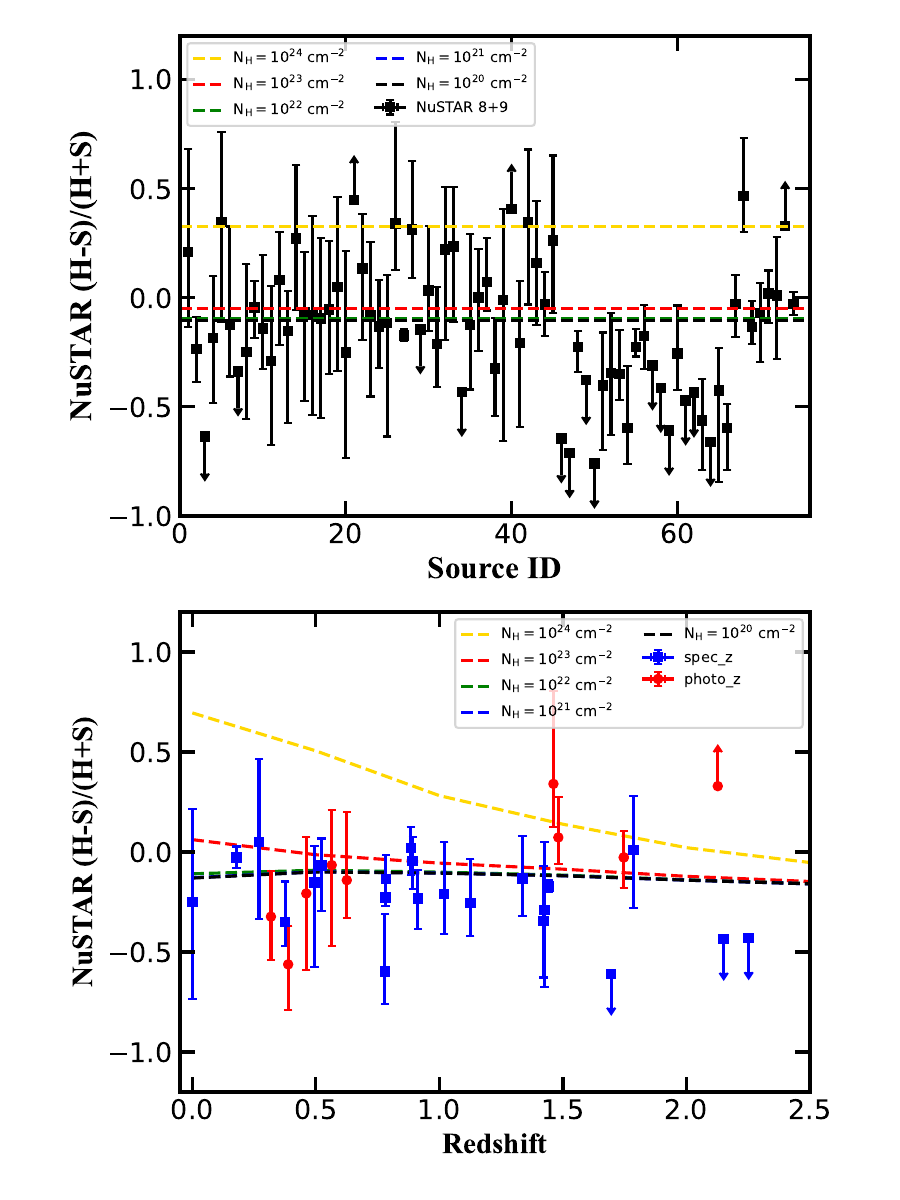}
    \caption{The hardness ratios of the 75 \nus sources detected in cycles 8+9. S and H represent the \thei and \eitw bands, respectively. The horizontal dashed lines indicate the likely column density associated with that hardness ratio assuming the median redshift of our sample, $z = $0.885. \textbf{Bottom panel:} The HRs vs $z$ for the 31 \nus sources with a known redshift. Neither of the two sources that fall in the CT regime (Nus$_{89}$\_ID = 27 and Nus$_{89}$\_ID = 74) were detected in the cycles 5+6 survey.}
    \label{fig:nus_hr}
\end{figure*}

Hardness ratios (HRs) are a measure of the difference in counts between two different energy bands and are very effective in identifying obscured AGN. They are calculated using the formula (H-S)/(H+S) where H is the counts in the hard band and S is the counts in the soft band. As obscuration increases, the soft X-rays are more suppressed than the hard X-rays, thus increasing the HR value towards 1. \\
\indent Figures \ref{fig:xmm_hr} and \ref{fig:nus_hr} display simulated HR curves for different column densities. We used a physically-motivated model that models the line-of-sight component with an absorbed power-law, the reflection component with \texttt{MYTorus} \citep{MYTorus2009}, and the scattered emission with a fractional power law. This was modeled in XSPEC using the equation \texttt{phabs$\times$(zphabs$\times$powerlw+MYTorus+\newline constant$\times$powerlw)}, where \texttt{phabs} accounts for the galactic absorption. For both power laws in \texttt{powerlw} and \texttt{borus02}, we used $\Gamma$=1.80. Using the properties found in \cite{Zhao2021a}, we set the average torus column density $N_{\rm H,Tor}$ = 1.4 $\times$ 10$^{24}$ cm$^{-2}$, the covering factor $f_c$ = 0.67, and the inclination angle $\theta_{inc}$ = 60$\degr$. As per \cite{Ricci2017}, we used a \texttt{constant} = 1\% to account for the scattering fraction. \\
\indent We calculated the \xmm HRs using the bands S = \zetw and H = 2$-$10\,keV. The HR distribution for all 274 cycles 8+9 \xmm sources are plotted in the top panel of Figure \ref{fig:xmm_hr}. Based on these results, 55\% of the 274 sources were found to be obscured ($N_{\rm H} \geq$ 10$^{22}$ cm$^{-2}$). This figure predicts column density assuming all sources have a redshift $z = 0$. The bottom panel of Figure \ref{fig:xmm_hr} shows the predicted column density for the 92 XMM sources with a known redshift. Of the 92, $\sim$40\% have an HR classifying them as obscured AGN. This is a similar rate to the cycles 5+6 \xmm survey, which found 38\% of sources to be obscured based on their hardness ratios. \\
\indent Figure \ref{fig:nus_hr} shows the same two plots for the 75 \nus cycles 8+9 sources. For \textit{NuSTAR}, we used \thei as the soft band S and \eitw as the hard band H. In line with \citetalias{Zhao2021} and \citetalias{Zhao2024}, these HRs were calculated using the Bayesian Estimation of Hardness Ratios \citep[BEHR;][]{Park2006}. We implemented BEHR because it is capable of calculating hardness ratios even when sources are in the Poisson regime due to limited count statistics. The method used to calculate the 1$\sigma$ uncertainty was determined based on the net counts in each band. If the net counts from either band were below 15, we used the Gaussian-quadrature numerical integration method. If both net counts were above 15, we used the Gibbs sampler method. We also considered the differences in effective exposure times from each band.\\
\indent Due to its sensitivity above 10\,keV, \nus is better equipped than \xmm for accurately determining the column density of heavily obscured ($N_{\rm H} >$ 10$^{23}$ cm$^{-2}$) AGN. Out of the entire sample, 27 sources (36\%) were found to be heavily obscured. Limiting to only the 31 \nus sources with redshift values, 8 (26\%) are heavily obscured and 2 (6\%) are CT. We note that these numbers do not include sources whose upper error bar crosses over into the heavily obscured or CT region. Due to the additional photometric redshifts obtained from JWST, we can place our two CT-AGN candidates at $z = 1.46$ (Nus$_{89}$\_ID = 27) and $z = 2.12$ (Nus$_{89}$\_ID = 74). \\
\indent We note that the HRs from XMM and those from \nus may be different due to how they are calculated. As mentioned above, our models assume spectral shapes that are used to convert count rates into fluxes. When you change the input NH from unabsorbed to CT, the ECF changes by 70\% in the \zetw band and 20\% in the \eitw band. Moreover, the ECF can change by 5\% (smaller but not insignificant) when the photon index is changed from 1.80 to 1.40 or 2.20. It is for these reasons that a full spectral analysis of both the XMM and \nus data is needed in order to accurately characterize the column densities and spectral indices of the detected sources. This work done on the 60 sources detected by \nus in \citetalias{Zhao2024} has been 
performed by Creech et al. submitted. \\

\subsection{CT Fraction}
\begin{figure*}
    \centering
    \includegraphics[width=0.5\linewidth]{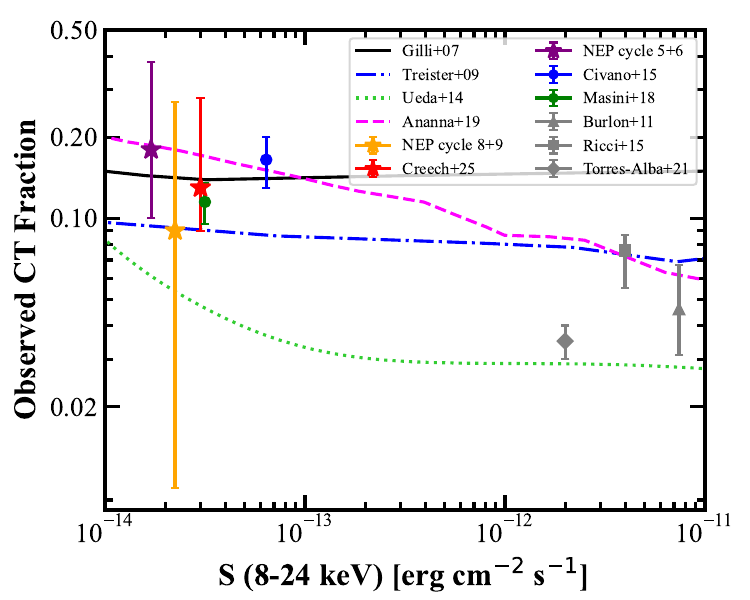}
    \caption{The CT-AGN fraction as a function of the \eitw band flux sensitivity from this work (orange star), \citetalias{Zhao2024} (purple star), and Creech et al. submitted (red star). Multiple population synthesis models are plotted as lines: the black solid line \citep{Gilli2007}, the blue dash-dotted line \citep{Treister2009}, the green dotted line \citep{Ueda2014}, and the magenta dashed line \citep{Ananna2019}. The blue and green circular points are from the COSMOS \citep{Civano2015} and UDS \citep{Masini2018a} fields. Finally, there are three grey points that display different \textit{Swift}-BAT data points: \cite{Burlon2011} (triangle), \cite{Ricci2015} (square), and \cite{Torres-Alba2021} (diamond). }
    \label{fig:ct_frac}
\end{figure*}

Because hard X-rays are less susceptible to obscuration, surveys above 10\,keV are well equipped to calculate the fraction of CT-AGN. As can be seen in the bottom panel of Figure \ref{fig:nus_hr}, there are two sources with known redshifts (Nus$_{89}$\_ID = 27 and 74) expected to be CT based on their \nus hardness ratios. Two more sources without redshift values (Nus$_{89}$\_ID = 22 and 41) are CT candidates as their HR is greater than the CT threshold (HR $>$ 0.736) for sources at $z = 0$. %This sets the CT lower limit for this cycle as 4 / 75 = 0.05. 
Now, if we include sources with HRs greater than the CT threshold (HR $>$ 0.325) for the median redshift in our sample ($z$ = 0.885), this adds three more sources (Nus$_{89}$\_ID = 6, 43, and 69). Thus, the estimate of the CT fraction in this survey is: 7 / 75 = 0.09. To find our lower limit, we look at how many of these seven sources have their lower uncertainty value greater than the CT threshold HR $>$ 0.736 (if the redshift is unknown) or greater than the yellow curve in Figure 20, bottom (if the redshift is known). Only one of the seven sources is above this threshold, thus the lower limit is 1. To find our upper limit, this includes all sources whose HR upper uncertainty value is greater than 0.325. This adds another 13 sources, making the CT-fraction upper limit: 20 / 75 = 0.27. Thus, the CT fraction for the \nus cycles 8+9 survey is $9^{+18}_{-8}$\%. This is lower (however still consistent within uncertainties) than the \nus cycles 5+6 survey which found a fraction of $18^{+20}_{-8}$\%. One potential explanation for this is that the cycles 5+6 survey was deeper, and thus more capable of detecting very faint CT-AGN. As stated in Section  \ref{sec:hard_ratios}, a detailed spectral analysis is needed to confirm this CT fraction, which should be considered an estimate when using hardness ratios alone. Creech et al. submitted performed this spectral analysis on the 60 sources from \citetalias{Zhao2024} and found a CT-fraction of $13^{+15}_{-4}$\%, agreeing within uncertainties with both the value presented in this work and \citetalias{Zhao2024}. \\
\indent Figure \ref{fig:ct_frac} displays the hardness ratio CT-fraction from cycles 8+9 (orange star) and cycles 5+6 (purple star), and the spectroscopically derived CT-fraction from Creech et al. submitted (red star), compared with other high-energy surveys and CXB population synthesis models. The other two \nus surveys found similar values to the \nus NEP surveys: the COSMOS field = 17\% $\pm$ 4\% \citep{Civano2015} and the UDS field = 11.5\% $\pm$ 2.0\% \citep{Masini2018a}. The grey points represent CT fractions obtained from \textit{Swift}-BAT surveys. The first two surveys, \cite{Burlon2011} and \cite{Ricci2015} found fractions of $\sim$4.6\% and $\sim$7.6\%. Using more BAT data, \cite{Torres-Alba2021} found a lower fraction of $\sim$3.5\%. However, this work noticed that the CT fraction is highly dependent on redshift. Limiting the sample to $z \leq 0.01$, the fraction rose to 20\% and for $z \leq 0.05$ it was 8\% (see Figure 3 in their paper). As previously mentioned, BAT is less sensitive than \textit{NuSTAR} and thus will detect even less CT-AGN as redshift increases. Finally, Figure \ref{fig:ct_frac} also displays four different CXB population synthesis models from \cite{Gilli2007}, \cite{Treister2009}, \cite{Ueda2014}, and \cite{Ananna2019}. The CT fraction from this work is consistent with all four within uncertainties, but most closely agrees with \cite{Treister2009}. This is different from the cycles 5+6 survey, which most closely agreed with \cite{Ananna2019}. This motivates our future works of determining the redshifts for more sources in our sample, performing detailed spectral analysis, and combining the data from all four cycles.

\section{Summary} \label{sec:conc_sec8}
This work presents the analysis of the cycles 8+9 \nus observations of the NEP TDF field that totaled $\sim$2.0\,Ms and covered $\sim$0.31\,deg$^2$. We analyzed 26 different observations taken between August of 2022 and May of 2024. This is the second deepest \nus extragalactic survey, following the \nus cycles 5+6 survey presented in \citetalias{Zhao2024}. The primary results of this survey are listed below.

\begin{enumerate}
    \item The combination of the cycles 8+9 \nus data yielded 75 detected sources above a 99\% reliability level. Out of the 60 sources detected in the cycles 5+6 \nus survey, 23 were detected in both. Therefore, 112 total sources have been detected by \nus during these four cycles of observations. \\
    \item The 8 combined \xmm observations from cycles 8+9 yielded 274 detections. Of the 286 sources detected by \xmm in cycle 6, 107 were detected in both surveys. Therefore, 453 unique sources have been detected by 
    \xmm during these three cycles. \\
    \item This \nus survey reached a 20\%-area flux of 2.2 $\times$ 10$^{-14}$ erg cm$^{-2}$ s$^{-1}$ in the 8$-$24\,keV band. Additionally, the \xmm survey reached a 20\%-area flux of 4.2 $\times$ 10$^{-15}$ erg cm$^{-2}$ s$^{-1}$ in the \twte band. \\
    \item This work found the slope of the \eitw \logn curve to be $\alpha_{89} = 1.13 \pm 0.46$ and the cycles 5+6 slope to be $\alpha_{56} = 1.24 \pm 0.41$. This is flatter than other works, but consistent with the Euclidean value of $\alpha = 1.50$. When the brightest source in our sample, the blazar, is removed, these values steepen ($\alpha_{89} = 1.23 \pm 0.50$ and $\alpha_{56} = 1.36 \pm 0.46$) and become more consistent with other works ($\alpha \sim 1.7$).
    \item Of the 75 \nus detected sources in this survey, 48 of them were found to have XMM counterparts from the 8+9 data. An additional 5 have XMM counterparts from the cycle 6 data. Moreover, three more sources have a \cha counterpart. Therefore, 56 / 75 (75\%) of the \nus sources have a soft X-ray counterpart. Moreover, we found \cha counterparts for 130 of the 274 XMM sources. \\ 
    \item We searched the multiwavelength catalogs of the NEP to find counterparts for the XMM sources in the radio (VLA), infrared (JWST, MMIRS, WISE), and optical (HST, HSC, SDSS). In total, 221 of the 274 (81\%) have at least one lower-energy counterpart. Additionally, 54 out of the 75 \nus sources from cycles 8+9 have at least one non-X-ray counterpart. This yields a completeness of 72\% for \nus non-X-ray counterparts. \\
    \item Approximately 55\% of the XMM sources were found to be obscured ($N_{\rm H} \geq$ 10$^{22}$ cm$^{-2}$) and 36\% of the \nus sources were found to be heavily obscured ($N_{\rm H} \geq$ 10$^{23}$ cm$^{-2}$) based on hardness ratios. We found a CT fraction of $9^{+18}_{-8}$\%. This is lower than the fraction found in \citetalias{Zhao2024} but still consistent within uncertainties. \\

\end{enumerate}

\indent The spectral analysis of the 60 \nus sources from \citetalias{Zhao2024} is reported in Creech et al. (submitted) with the goal of obtaining accurate spectral properties (column densities and photon indices). While the last work includes also the data used in this paper, the 52 sources reported here and not in \citetalias{Zhao2024} were not part of their analysis and will be included in a future work. Additionally, as time domain was the primary focus of the NEP field, a future work will study all the X-ray data of the NEP to obtain more in-depth results on the source variability. Additionally, we plan to continue previous optical campaigns with Hectospec, Binospec, and GMOS-N to obtain more spectroscopic redshifts of the X-ray detected NEP TDF sample.

\section{Acknowledgments}
The authors thank the \nus and \xmm teams for their efforts to schedule simultaneous observations in each epoch of this survey. Having simultaneous observations removed the need for flux calibrations between observatories, streamlining the analysis of this survey. \\
\indent The authors thank the anonymous referee for their helpful comments which greatly improved our paper. \\
The material is based upon work supported by NASA under award number 80GSFC24M0006. \\
\indent We dedicate this paper to the memory of our dear PEARLS colleague Mario Nonino, who was a gifted and dedicated scientist, and a generous person. This work is based on observations made with the NASA/ESA/CSA James Webb Space Telescope. The data were obtained from the Mikulski Archive for Space Telescopes at the Space Telescope Science Institute, which is operated by the
Association of Universities for Research in Astronomy, Inc., under NASA contract NAS 5-03127 for JWST. These observations are associated with JWST programs 1176 and 2738. \\
\indent RAW, SHC, and RAJ acknowledge support from NASA JWST Interdisciplinary Scientist grants NAG5-12460, NNX14AN10G and 80NSSC18K0200 from GSFC. Work by CJC acknowledges support from the European Research Council (ERC) Advanced Investigator Grant EPOCHS (788113). BLF thanks the Berkeley Center for Theoretical Physics for their hospitality during the writing of this paper.
CNAW acknowledges funding from the JWST/NIRCam contract NASS-0215 to the University of Arizona. We thank the CANUCS team for generously providing early access to their proprietary data of MACS0416. \\
\indent We also acknowledge the indigenous peoples of Arizona, including the Akimel O'odham (Pima) and Pee Posh (Maricopa) Indian Communities, whose care and keeping of the land has enabled us to be at ASU's Tempe campus in the Salt River Valley, where much of our work was conducted.

\section{Data Availability}
The \nus and \xmm catalogs will be available for download on VizieR following the publication of this paper.

\section{Appendix}

\begin{figure}
    %\centering
    \hspace{-1.8cm}
    \includegraphics[scale=0.47, clip=true,trim=75 0 0 0]{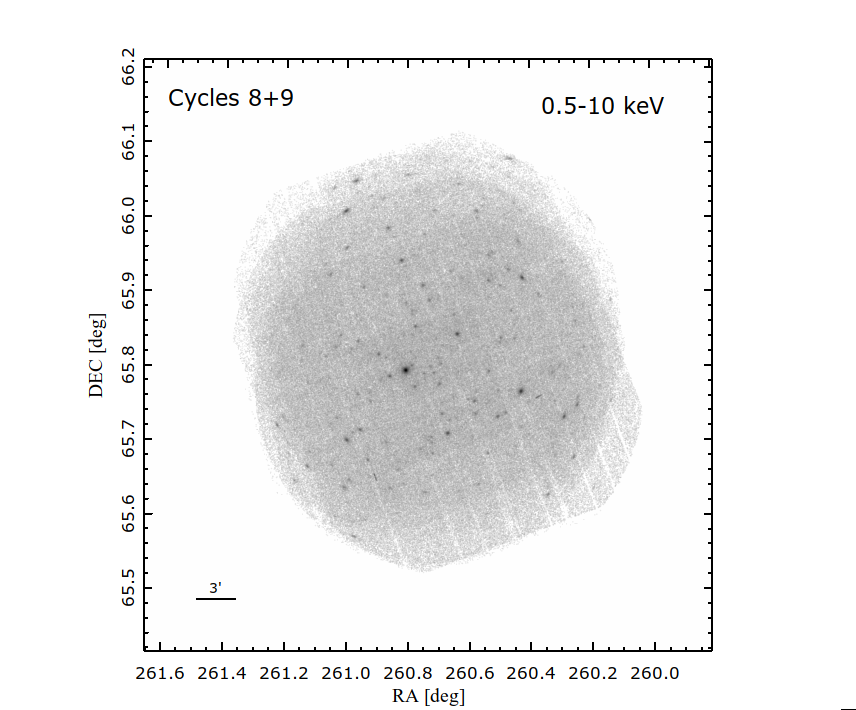}
    \caption{The combined mosaic of the eight \xmm observations from cycles 8+9 in the 0.5$-$10\,keV band. No source regions are plotted to make the sources more visible. }
    \label{fig:xmm_05-10_no_regs}
\end{figure}

\begingroup
\renewcommand*{\arraystretch}{1.1}
\begin{table*}
\scriptsize
\caption{\nus 99\% Reliability Source Catalog Description}
\centering
\label{Table:nus_cat_info}
  \begin{tabular}{ll}
   \hline
   \hline
 	Col. & Description \\
	\hline
    1 & \nus source ID used in this paper\\
    2 & Source name (use “NuSTAR JHHMMSS+DDMM.m”) \\
    3-4 & X-ray coordinates (J2000) of the source in whichever energy band has the highest DET\_ML \\
    5 & \thtw band deblended DET\_ML (-99 if the source is not detected in a given band) \\
    6 & \thtw band vignetting-corrected exposure time in kiloseconds at the position of the source \\
    7 & \thtw band total counts (source + background) in a 20$\arcsec$ radius aperture \\
    8 & \thtw band deblended background counts in a 20$\arcsec$ radius aperture (-99 if the source is not detected in a given band) \\
    9 & \thtw band not deblended background counts in a 20$\arcsec$ radius aperture \\
    10 & \thtw band net counts (deblended if detected and above DET\_ML threshold or 90\% confidence upper limit if \\
    & undetected or detected but below DET\_ML threshold) in a 20$\arcsec$ radius aperture \\
    11-12 & \thtw band 1$\sigma$ positive/negative net counts uncertainty (-99 for upper limits) \\
    13 & \thtw band count rate (90\% confidence upper limit if not detected or detected but below the threshold) in \\
    & a 20$\arcsec$ radius aperture \\
    14 & \thtw band aperture-corrected flux (erg cm$^{-2}$ s$^{-1}$; 90\% confidence upper limit if below 99\% confidence threshold) \\
    15-16 & \thtw band positive/negative flux uncertainties (-99 for upper limits) \\
    17-28 & Same as columns (5)–(16) but for \thei \\
    29-40 & Same as columns (5)–(16) but for \eitw \\
    41-52 & Same as columns (5)–(16) but for \eisi \\
    53-64 & Same as columns (5)–(16) but for \sitw \\
    65 & HR computed using BEHR \\
    66-67 & Lower/upper limit of HR \\
    68 & Source ID in the \citetalias{Zhao2024} NuSTAR cycles 5+6 catalog (-99 for nondetection in the cycles 5+6 catalog) \\
    69 & \xmm source ID from the XMM-Newton catalog (-1 if nondetection) \\
    70-71 & \xmm coordinates of the associated source (-1 if no XMM-Newton counterpart) \\
    72 & \nus to \xmm counterpart position separation in arcseconds \\
    73 & \thei flux converted from \xmm \twte flux (erg cm$^{-2}$ s$^{-1}$; 90\% confidence upper limit if \texttt{mlmin} $<$ 6) \\
    74 & \thei \xmm flux 1$\sigma$ uncertainty (-99 for upper limit)\\
    75 & Flag for NuSTAR counterparts (S, P, Sec, or C if the XMM source is the single, primary, secondary, or \\
    & confusing counterpart of the NuSTAR source, respectively) \\
    76 & Flag for ancillary class (S for secure, A for ambiguous, and U for unidentified) \\
    77-78 & Ancillary coordinates of the associated source (-99 if no detection) \\
    79 & Optical (HSC) ID (-99 if no detection) \\
    80 & Optical (HSC) i-band AB magnitude (-99 if no detection) \\
    81 & MMIRS ID (-99 if no detection) \\
    82 & MMIRS J-band AB magnitude (-99 if no detection) \\
    83 & WISE ID (-99 if no detection) \\
    84 & WISE W1-band AB magnitude (-99 if no detection) \\
    85 & VLA 3 GHz counterpart ID from \cite{Hyun2023} \\
    86 & VLA 3 GHz flux density in $\mu$Jy \citep{Hyun2023} \\
    87 & HST counterpart ID from (-99 if no detection) \citep[][Jansen et al. in preparation]{Obrien2024} \\
    88 & HST F606W AB magnitude (-99 if no detection) \citep[][Jansen et al. in preparation]{Obrien2024} \\
    89 & JWST counterpart ID (-99 if no detection) \\
    90 & JWST F444W AB magnitude (-99 if no detection) \\
    91 & \cha counterpart ID (Maksym et al. in prep) \\
    92 & \cha net count rate in the 0.5$-$7\,keV band (Maksym et al. in prep) \\
    93 & Spectroscopic redshift of the associated source \\
    94 & Photometric redshift of the associated source \\
    95 & Spectroscopic classification (Q for quasars, G for galaxies, S for stars, N/A if no measurement); galaxies are \\
    & defined as objects without broad emission lines and therefore include type 2 AGN \\
    96 & Luminosity distance in Mpc (-99 if no redshift measurement) \\
    97 & 10--40\,keV band rest-frame luminosity (-99 if no redshift measurement) \\
    98-99 & 10--40\,keV band 1$\sigma$ positive/negative rest-frame luminosity uncertainty (erg s$^{-1}$; -99 if no redshift measurement) \\
    
    %\fix{Things to add:} & \\
    %& Chandra ID / coordinates / ct rate (flux?) \\
    %& ID for all MW counterparts \\
    %& XMM 6 ID \\

   \hline   	
	
\end{tabular}
\end{table*}
\endgroup

\begingroup
\renewcommand*{\arraystretch}{1.1}
\begin{table*}
\caption{\xmm Source Catalog Description}
\centering
\label{Table:xmm_cat_info}
  \begin{tabular}{ll}
   \hline
   \hline
 	Col. & Description \\
	\hline
    1 & \xmm source ID used in this paper\\
    2 & \xmm source name (use “TDFXMM JHHMMSS+DDMM.m”) \\
    3-4 & X-ray coordinates (J2000) of the source in whichever energy band has the highest DET\_ML \\
    5 & \zetw band DET\_ML (-99 if the source is not detected in this band) \\
    6 & \zetw band vignetting-corrected exposure time in kiloseconds at the position of the source \\
    7 & \zetw band net counts of the source (90\% confidence upper limit if \texttt{mlmin} $<$ 6) \\
    8 & \zetw band net counts 1$\sigma$ uncertainty (-99 for upper limits) \\
    9 & \zetw band flux (erg cm$^{-2}$ s$^{-1}$; 90\% confidence upper limit if \texttt{mlmin} $<$ 6) \\
    10 & \zetw band flux 1$\sigma$ error (-99 for upper limits) \\
    11-16 & Same as columns (5)–(16) but for \twte \\
    17 & HR (90\% confidence upper or lower limits if not constrained) \\
    18 & HR 1$\sigma$ uncertainty (-99 for upper limits and 99 for lower limits) \\
    19 & \nus source ID from the \nus cycles 8+9 catalog (-1 if nondetection) \\
    20 & Flag for \nus cycles 8+9 counterparts (S, P, Sec, or C if the XMM source is the single, primary, \\
    & secondary, or confused counterpart of the NuSTAR source, respectively) \\
    21 & \nus source ID from the NuSTAR cycles 5+6 catalog (-1 if nondetection) \\
    22 & Flag for \nus cycles 5+6 counterparts (S, P, Sec, or C if the XMM source is the single, primary, \\
    & secondary, or confused counterpart of the NuSTAR source, respectively) \\
    23 & Flag for ancillary class (S for secure, A for ambiguous, or U for unidentified) \\
    24-25 & Ancillary coordinates of the associated source (-99 if no detection) \\
    26 & Optical (HSC) ID (-99 if no detection) \\
    27 & Optical (HSC) i-band AB magnitude (-99 if no detection) \\
    28 & Flag for SDSS detection (1 if SDSS has detection, -1 if SDSS has no detection) \\
    29 & MMIRS ID (-99 if no detection) \\
    30 & MMIRS J AB magnitude (-99 if no detection) \\
    31 & WISE ID (-99 if no detection) \\
    32 & WISE W1 AB magnitude (-99 if no detection) \\
    33 & VLA 3 GHz counterpart ID from \cite{Hyun2023} \\
    34 & VLA 3 GHz flux density in $\mu$Jy \citep{Hyun2023} \\
    35 & HST ID (-99 if no detection) \\
    36 & HST F606W AB magnitude (-99 if no detection) \citep[][Jansen et al. in preparation]{Obrien2024} \\
    37 & JWST ID (-99 if no detection) \\
    38 & JWST F444W AB magnitude (-99 if no detection) \\
    39 & \cha counterpart ID (Maksym et al. in prep) \\
    40 & \cha net count rate in the 0.5$-$7\,keV band (Maksym et al. in prep) \\
    41 & Spectroscopic redshift of the associated source \\
    42 & Photometric redshift of the associated source \\
    43 & Spectroscopic classification (Q for quasars, G for galaxies, S for stars, N/A if no measurement); galaxies are \\
    44 & Luminosity distance in Mpc (-99 if no redshift measurement) \\
    45 & \zetw band rest-frame luminosity before correcting for absorption assuming a photon index of $\Gamma$ = 1.40 \\
    & (erg s$^{-1}$; -99 if not detected in the \zetw band) \\
    46 & \zetw band rest-frame luminosity 1$\sigma$ uncertainty \\
    47 & \twte band rest-frame luminosity before correcting for absorption assuming a photon index of $\Gamma$ = 1.80 \\
    & (erg s$^{-1}$; -99 if not detected in the \twte band) \\
    48 & \twte band rest-frame luminosity 1$\sigma$ uncertainty\\
   \hline   	
	
\end{tabular}
\end{table*}
\endgroup

\bibliographystyle{aasjournal}
\bibliography{bibliography}

\end{document}